\documentclass[twoside,12pt]{article}
\usepackage{epsfig}
\usepackage{amssymb}
\usepackage{amsmath,ulem,cancel}
\usepackage[pdftex,a4paper,colorlinks=true,citecolor=blue,urlcolor=blue]{hyperref}
\usepackage[numbers,sort&compress]{natbib}

\newcommand{\be}{\begin{equation}}
\newcommand{\ee}{\end{equation}}
\newcommand{\bea}{\begin{eqnarray}}
\newcommand{\eea}{\end{eqnarray}}

\begin{document}

\title{ \vspace{1cm} LUNA: Status and Prospects}
\author{C.\ Broggini,$^{1}$ D.\ Bemmerer,$^{2}$ A.\ Caciolli,$^3$ D.\ Trezzi,$^4$ \\
\\
$^1$INFN, Sezione di Padova, Italy\\
$^2$Helmholtz-Zentrum Dresden-Rossendorf (HZDR), Dresden, Germany\\
$^3$Dipartimento di Fisica e Astronomia dell'Universit\`a and INFN, Padova,
Italy\\
$^4$Universit\`a degli Studi di Milano and INFN, Milano,
Italy\\
}
\maketitle
\begin{abstract}
The essential ingredients of nuclear astrophysics are the thermonuclear reactions
which shape the life and death of stars and which are responsible for the synthesis of the chemical elements in the Universe.
Deep underground in the Gran Sasso Laboratory the cross sections of the key reactions responsible for the hydrogen burning in stars
have been measured  with two accelerators of 50 and 400 kV voltage right down to the energies of astrophysical interest.
As a matter of fact, the main advantage of the underground laboratory is the reduction of the background. Such a reduction has allowed, for the first time, to measure relevant cross sections at the Gamow energy.
The qualifying features of underground nuclear
astrophysics are exhaustively reviewed before discussing the current LUNA program which is mainly devoted to the study of the Big-Bang nucleosynthesis
and of the synthesis of the light elements in AGB stars and classical novae.
The main results obtained during the study of reactions relevant to the Sun are also reviewed and their influence on our understanding of the properties of the neutrino, of the Sun and of the Universe itself is discussed. Finally, the future of LUNA during the next decade is outlined. It will be mainly focused on the study of the nuclear burning stages after hydrogen burning:
helium and carbon burning. All this will be accomplished thanks to a new 3.5 MV accelerator able to deliver high current beams of proton, helium and carbon which will start
running under Gran Sasso in 2019. In particular, we will discuss the first phase of the scientific case of the 3.5 MV accelerator focused on the study of
$^{12}$C+$^{12}$C and of the two reactions which  generate free neutrons inside stars: $^{13}$C($\alpha$,n)$^{16}$O and $^{22}$Ne($\alpha$,n)$^{25}$Mg.
\end{abstract}

\section{Introduction}\label{sec:intro}

\begin{figure}[htb]
\centerline{%
\includegraphics[width=0.5\textwidth]{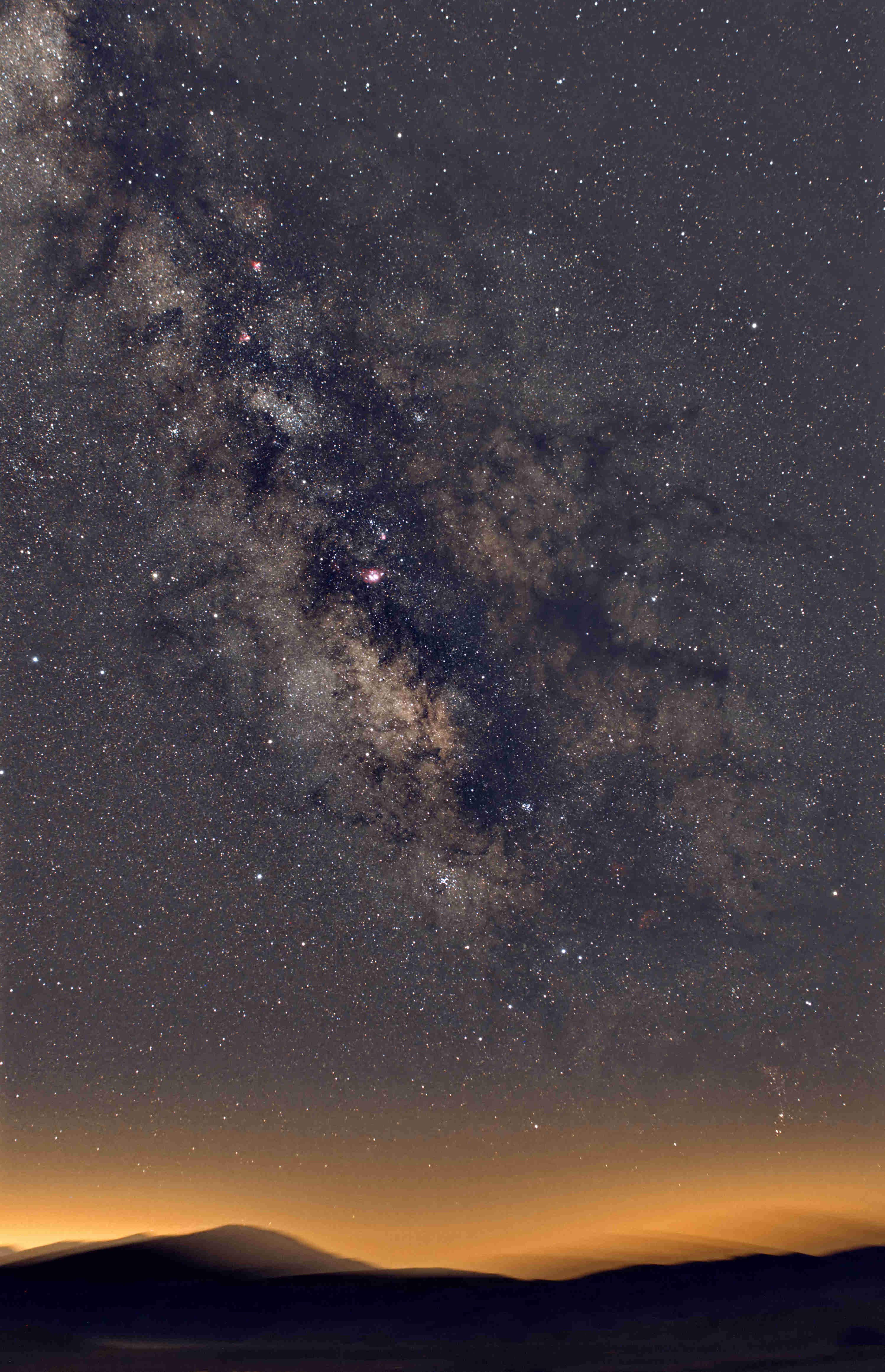}}
\caption{\label{fig:MilkyWay_GranSasso}
The Milky Way over Gran Sasso on a moonless night.
Under Gran Sasso LUNA studies the thermonuclear reactions which supply stars with the power to shine. }
\end{figure}

Gravity triggers the birth of a star but the loss of gravitational energy alone is not enough to supply the  energy radiated away during the entire lifetime of a star
(Fig.\ref{fig:MilkyWay_GranSasso}).
A different and fundamental source of energy must necessarily be assumed \cite{Eddington20-Nature}. This source  is given by the binding energy of atomic nuclei, which can be released in particular when converting protons (zero binding energy) up to medium-mass nuclei such as $^{56}$Fe (binding energy per nucleon $E_{\rm B}/A$ = 8.6 MeV). In order to properly understand stellar structure and evolution, it is thus necessary to understand how light nuclei are converted to heavier ones: It is necessary to study nuclear reactions.

In addition to energy production and its effects on stellar structure and evolution, nuclear reactions also affect the production of neutrinos in stars. Our Sun is an example of a star with a well-studied neutrino spectrum \cite{Davis03-NobelLecture,McDonald16-NobelLecture}. Finally, nuclear reactions in stars synthesize chemical elements. All but the lightest five elements (hydrogen to boron) are predominantly produced in stars \cite{Clayton03-Book}, and for example the crucial ratio of carbon to oxygen abundances is controlled by stellar helium burning.

The key point of a nuclear reaction is the value of its cross section at the
energy at which the reaction takes place.
The extremely low value of the cross section at the stellar energies, ranging
from pico to femto-barn and even below, has always
prevented its measurement in a laboratory at the Earth's surface,
where the signal to background ratio is too small mainly because
of cosmic ray interactions. Instead, the observed energy dependence of the
cross-section at high energies is extrapolated to the low energy
region, leading to substantial uncertainties.
In particular, the reaction mechanism might change, or there might be the
contribution of narrow or of large sub-threshold resonances which cannot
be accounted for by the extrapolation, but which could completely
dominate the reaction rate at the stellar energies.

\begin{figure}[htb]
\centerline{%
\includegraphics[width=0.8\textwidth]{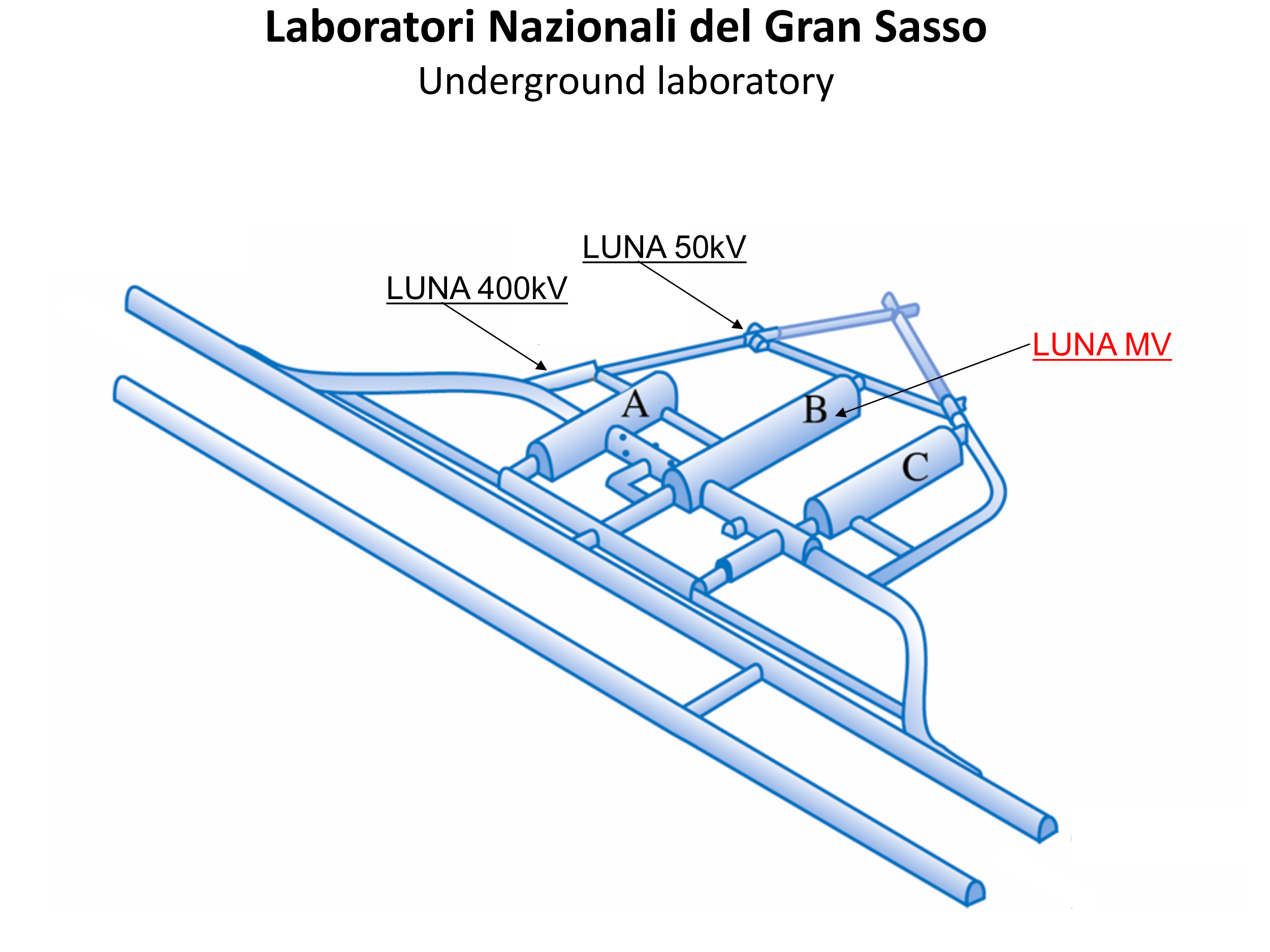}}
\caption{\label{fig:PiantaLNGS}
The map of the underground Gran Sasso Laboratory, with the location of the different LUNA accelerators.}
\end{figure}

In order to explore the low energy
domain of nuclear astrophysics
LUNA (Laboratory for Underground Nuclear Astrophysics) started its activity in 1991
as a pilot project
with a 50 kV electrostatic accelerator \cite{Greife94-NIMA}
installed inside the laboratory under Gran Sasso
 (Fig.\ref{fig:PiantaLNGS}).

LUNA still remains the only experiment in the world running an accelerator deep underground (at the moment a 400 kV accelerator producing hydrogen and helium beams \cite{Formicola03-NIMA}, LUNA 400)
but its achievements have triggered two similar facilities which are now going to start in the Republic of
China \cite{Liu16-EPJWOC} and in the United States \cite{Robertson16-EPJWOC}. A project for nuclear astrophysics is also active at the Canfranc Laboratory in Spain
\cite{CANFRANC}.
As a matter of fact, the extremely low laboratory background under Gran Sasso has allowed for the first time nuclear physics experiments with very small count rates,
down to a couple of events per month. This way, the important reactions responsible for the hydrogen burning in the Sun could be studied down to the relevant stellar energies.

At the end of the solar phase, i.e. study of reactions relevant to the Sun, LUNA started a rich program devoted to the study of the Big Bang Nucleosynthesis (BBN) and of the synthesis of the
elements through the CNO, Ne-Na and Mg-Al cycles. The motivation here is to reproduce the abundance of the light elements and
to identify the production site in stellar scenarios different from the Sun: hydrogen burning at the higher energies corresponding to the hydrogen shell of Asymptotic Giant Branch
(AGB) stars or to the explosive phase of classical novae.

Finally, time has now come to face the next steps beyond hydrogen burning: helium and carbon burning. Their study will be performed with the new 3.5 MV single-ended accelerator
which is going to be installed in Gran Sasso in 2018: LUNA MV \cite{LUNA15-AR,Aliotta16-EPJA}. The accelerator will provide hydrogen, helium and carbon (also double ionized) beams and it will
be devoted to the study of those key reactions of the helium and carbon burning which are determining and shaping both the evolution of massive stars towards their final
fate and the synthesis of most of the elements in the Universe.

In this review the main features of thermonuclear reactions at very low energy, the characteristics of the background attainable in Gran Sasso and the experimental apparata employed by LUNA
will first be described. Then, an overview of the improvements provided by LUNA to the picture of BBN and of hydrogen burning in stars, in particular the Sun,  will be given.
Finally, the next steps of nuclear astrophysics under Gran Sasso, mainly devoted to the study of the helium and carbon burning, will be outlined. In particular, we will discuss the reactions which are planned to be studied during the first phase
with the new accelerator:
$^{12}$C+$^{12}$C and the two reactions which  generate inside stars the neutrons responsible for the s-process, $^{13}$C($\alpha$,n)$^{16}$O and $^{22}$Ne($\alpha$,n)$^{25}$Mg.

\section{Thermonuclear reactions}\label{sec:thermo}

For most stellar scenarios, the characteristic time for changes in the system is much larger than the time for elastic collisions between the atomic components of the star. Thus the concept of temperature is well-defined, giving rise to the concept of thermonuclear reactions: Both partners in a nuclear reaction obey a Maxwell-Boltzmann velocity distribution. Assuming equal temperatures for the two reaction partners, the  distributions can be united into one called $\phi(v)$ for the relative velocity $v$  in the reaction's center of mass
(or P(E) for the energy in the center of mass), and the so-called thermonuclear reaction rate $\langle \sigma v \rangle$ can be defined:
%
\begin{eqnarray}
\langle \sigma v \rangle & \stackrel{\rm Def}{=} & \int_0^\infty \sigma(v) \; v \; \phi(v)  \; dv = \\
 & = & \int_0^\infty \sigma(E)  \; v(E)  \; P(E)  \; dE = \nonumber \\
 & = & \sqrt{\frac{8}{\pi\mu}} (k_{\rm B}T)^{-3/2} \int_0^\infty \sigma(E)  \; E  \; \exp[-E/k_{\rm B}T]  \; dE \label{eq:reaction_rate}
\end{eqnarray}
where 
$k_{\rm B}$ is Boltzmann's constant, $T$ the temperature, $\mu=m_1m_2/(m_1+m_2)$ the reduced mass of the two partners $m_1$ and $m_2$, and $\sigma (E)$ the nuclear reaction cross section as a function of center-of-mass energy $E$ \cite{Clayton84-Book,Iliadis07-Book}.

The thermonuclear reaction rate depends on the cross section $\sigma(E)$, which is different for each nuclear reaction, and on its energy dependence. Typical stellar temperatures $T$ $\sim$ 10$^7$-10$^9$\,K ($T_9$ = 0.01-1, expressed as $T_9$=$T$/1\,GK) correspond to peak energies of the Maxwell-Boltzmann distribution of $k_{\rm B}T$ $\sim$ 0.9-90\,keV. For reactions involving charged particles, these energies are well below the $\sim$ MeV high Coulomb barrier given by the electrostatic repulsion. The reaction cross section $\sigma(E)$ can be parameterized by the so-called astrophysical S-factor $S(E)$, using the relation
%
\begin{equation}\label{eq:sfactor}
\sigma(E) = \frac{S(E)}{E} \exp \left[-2 \pi \eta \right] = \frac{S(E)}{E} \exp \left[-2 \pi \frac{Z_1 Z_2 e^2}{\hbar} \sqrt{\frac{\mu}{2E}} \right]
\end{equation}
%
where the energy $E$ is in keV, $Z_{1,2}$ are the nuclear charges of the two reaction partners, $e$ is the elementary charge, and $\hbar$ the Planck's constant. The strongly energy dependent tunneling probability for the Coulomb barrier, given by the exponential term, and the geometrical factor $1/E$ are thus removed, and the purely nuclear physics term $S(E)$ remains.  In cases where the direct-capture process dominates, $S(E)$ depends very weakly on energy and lends itself to extrapolation e.g. using the R-matrix framework \cite{Descouvemont10-RPP}. In case of non-vanishing angular momentum, in addition to the Coulomb barrier also a centrifugal barrier must be taken into account, leading to an even steeper energy dependence of the cross section \cite{Iliadis07-Book}. However, even in those cases it is convenient to express the cross section in terms of $S(E)$, because when inserting eq.~\ref{eq:sfactor} into eq.~\ref{eq:reaction_rate} it is seen that $\langle \sigma v \rangle$ $\propto$ $S(E)$.

\begin{figure}[htb]
\centerline{%
\includegraphics[width=0.7\textwidth]{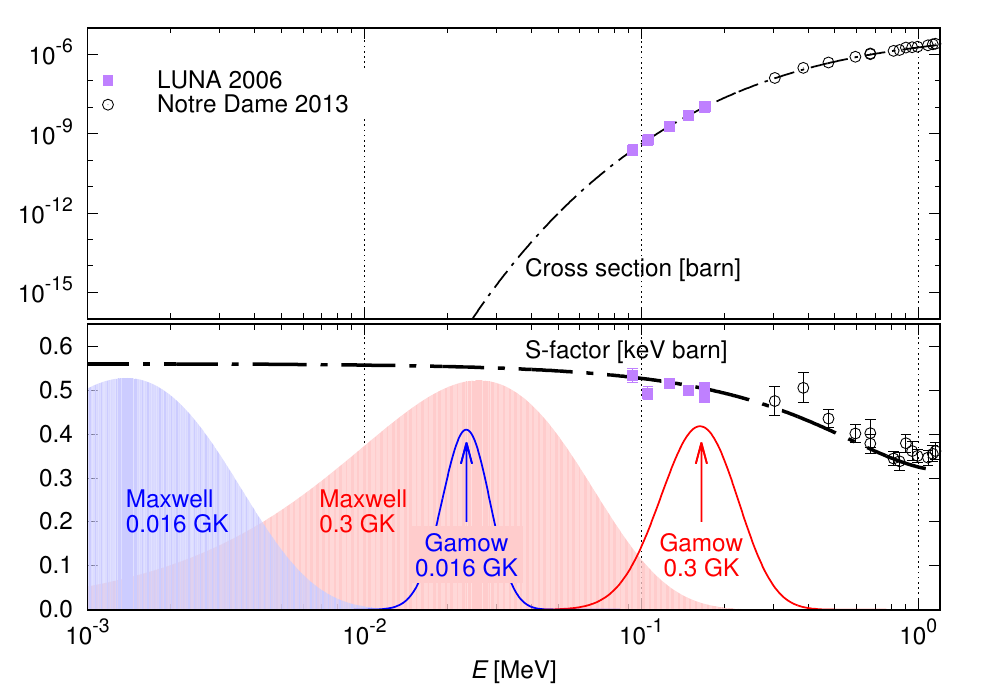}}
\caption{\label{fig:sfactor}
Using the $^3$He($\alpha$,$\gamma$)$^7$Be reaction as an example, the cross section (top panel) and the astrophysical S-factor (bottom panel) are plotted. For two temperatures typical of the Big Bang (0.3 GK) and of the Sun (0.016 GK), respectively, the Maxwell-Boltzmann distributions and Gamow peaks are added to the bottom panel. The Gamow energies are indicated by arrows. The experimental data are from LUNA \cite{Bemmerer06-PRL,Confortola07-PRC,Gyurky07-PRC} and from the lowest-energy surface-based experiment, Notre Dame \cite{Kontos13-PRC}.}
\end{figure}

The different energy behaviour of the three quantities $\sigma(E)$, $S(E)$, and $P(E)$ is plotted in Fig. \ref{fig:sfactor}, as well as a fourth quantity, the integrand in eq.~\ref{eq:reaction_rate}. This integrand is given by the product of the low-energy tail of $\sigma(E)$ and the high-energy tail of $P(E)$ and forms a peak. The shape of this peak is given by the product of the two exponential functions in eqns.~(\ref{eq:reaction_rate}, \ref{eq:sfactor}), $\exp[-E/k_{\rm B}T-b/\sqrt{E}]$, and indicates the energy range which contributes to the integrated reaction rate $\langle \sigma v \rangle$. It is frequently called the Gamow peak, and its maximum the Gamow energy. The Gamow peak may be approximated with a Gaussian \cite{Iliadis07-Book}, however with some deviations at low energies, and also for nuclear charges of $Z$ $\sim$ 20 and higher \cite{Rauscher10-PRC}.

\begin{figure}[htb]
\centerline{%
\includegraphics[scale=1]{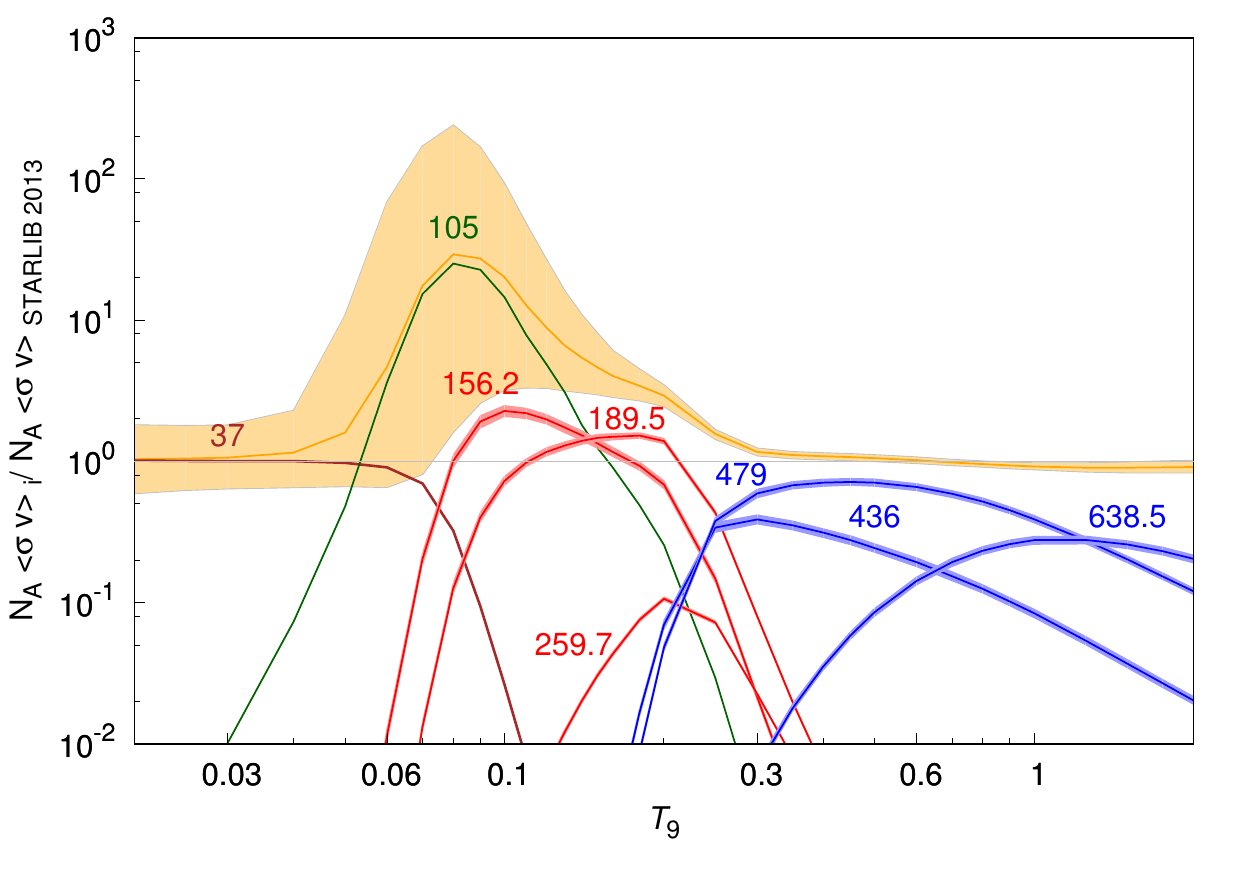}}
\caption{Astrophysical reaction rate for the $^{22}$Ne(p,$\gamma$)$^{23}$Na reaction \cite{Cavanna15-PRL,Depalo15-PRC,Depalo16-PRC}, normalized to the one of the latest compilation \cite{starlib}. The total rate is shown in yellow with its error band, and the contributions of selected individual resonances in color.}\label{fig:relative_rate}
\end{figure}

The scenario is different when considering nuclear resonances, which lead to a strong enhancement of $\sigma(E)$ at the resonance energy. For simplicity, here only a narrow, isolated resonance will be discussed, i.e. a resonance with an energetic width in the entrance channel that is much smaller than the distance to the nearest other resonance. It can be shown that for the usual Breit-Wigner \cite{BreitWigner36-PR} resonance shape, the reaction rate is given by
%
\begin{equation}\label{eq:reaction_rate_BW}
\langle \sigma v \rangle = \left(\frac{2\pi}{\mu k_{\rm B}T}\right)^{3/2}\hbar^2 \exp[-E_{\rm res}/k_{\rm B}T] \; \omega \gamma
\end{equation}
where $E_{\rm res}$ is the resonance energy and
\begin{equation}\label{eq:omegagamma}
\omega \gamma = \frac{2J + 1}{(2j_p+1)(2j_t+1)} \frac{\Gamma_{in}\Gamma_{ex}}{\Gamma}
\end{equation}
is the so-called resonance strength, which is, in turn, given by the product of a statistical factor calculated from the total angular momenta of the resonance, $J$, and of projectile and target nucleus $j_{p,t}$, and the ratio of entry and exit channel widths $\Gamma_{in,ex}$ and total width $\Gamma$ of the resonance. For a nuclear reaction dominated by one resonance, it is found that $\langle \sigma v \rangle$ $\propto$ $\omega\gamma$. For several well-separated resonances as in the recent example of the $^{22}$Ne(p,$\gamma$)$^{23}$Na reaction \cite{Cavanna15-PRL}, the contributions by each resonance can be summed, taking into account also the dependence on the resonance energy (Fig.\ref{fig:relative_rate}).

For both non-resonant and resonant nuclear reactions in the laboratory, the repulsive action of the positive nuclear charge is mitigated by the negatively charged atomic electrons. This results in a reduction of the Coulomb barrier height, and it has been shown that this effect, called the electron screening effect, may be parameterized with just one parameter, the so-called screening potential $U_e$. In the adiabatic limit, $U_e$ is simply given by the difference in the sum of electron binding energies in the incoming reaction channels and in the
combined atomic system which is formed \cite{Assenbaum87-ZPA}.

For the reactions studied at LUNA, the experimentally determined laboratory cross sections and resonance strengths have been corrected down by the screening enhancement factor
%
\begin{equation}\label{eq:screening}
\mathit{f} = \sigma_{\rm laboratory}(E) / \sigma_{\rm bare}(E) \sim \exp(\pi \eta U_e/E)
\end{equation}
%
with $\sigma_{\rm laboratory,\;bare}$ the cross sections in the laboratory and for a bare nucleus without atomic electrons, respectively, and $U_e$ given by its value at the adiabatic limit. However, it should be noted that for some cases of hydrogen implanted in metallic matrices enhanced screening has been found \cite{Raiola02-PLB,Czerski04-EPL}. Of course, the thermonuclear reaction rate thus computed for bare nuclei has to be subsequently corrected back for electron screening effects in the stellar plasma \cite{Salpeter54-AustralJP}.

Summarizing, in order to know the thermonuclear reaction rate $\langle \sigma v \rangle$ at the relevant stellar temperatures, for non-resonant nuclear reactions the astrophysical S-factor $S(E)$ must be known inside the Gamow peak, and for resonant nuclear reactions the energies and resonance strengths $\omega\gamma$ of resonances near the Gamow energy must be determined.

\section{Techniques for Underground Nuclear Astrophysics}\label{sec:underground}

The present section is devoted to the technical aspects which are important for an underground nuclear astrophysics experiment. The most important ingredient is the background, because it is radically different from the usual situation at the surface of the earth; therefore it is discussed first (sec.\ref{subsec:background}). Subsequently, the generation of the ion beams (sec.\ref{subsec:accelerators}), target assemblies (sec.\ref{subsec:targets}), and finally detectors (sec.\ref{subsec:detectors}) are discussed.

\subsection{Background}\label{subsec:background}

Typical LUNA experiments have very low signal counting rates; therefore it is important to have a low, and well-characterized, background. For the Big Bang and hydrogen burning scenarios hitherto studied at LUNA, the signal rate ranges from hundreds of counts per hour down to just a few counts per month, depending on the experiment.

The background seen in the detectors is classified here into two different categories: First, the environmental background, which includes the effects by cosmic rays and by the natural radioactivity in the experimental hall and in the setup, and, second, the beam induced background from parasitic reactions induced on contaminants in the setup.

\subsubsection{Environmental background at LUNA}

\begin{figure}[htb]
\centerline{%
\includegraphics[width=0.8\textwidth]{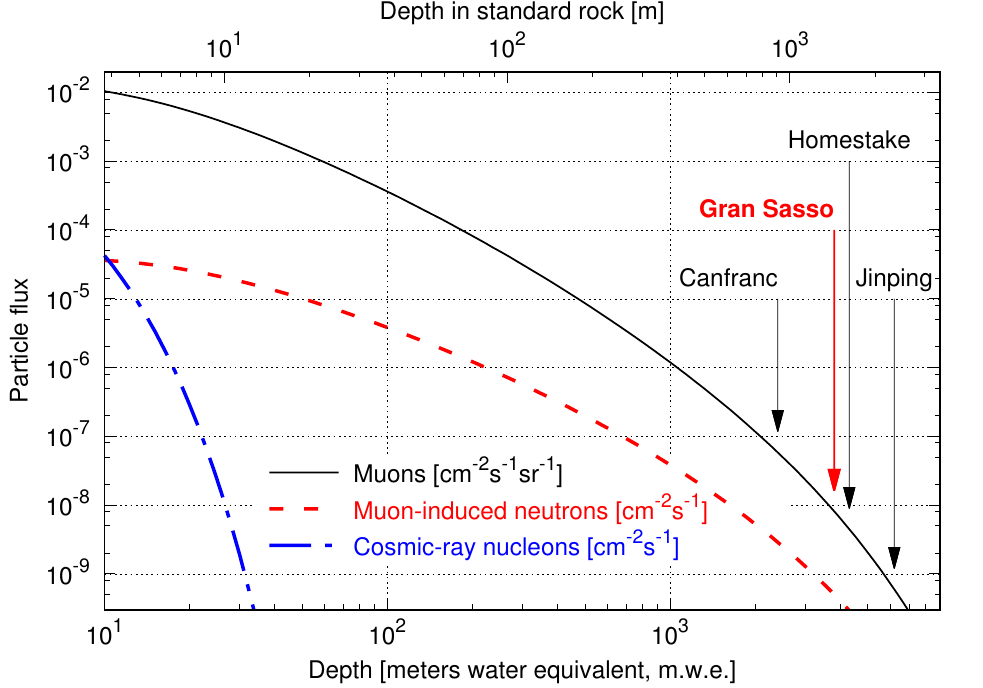}}
\caption{Muon, muon-induced neutron and cosmic-ray nucleon fluxes as function of depth. In addition to Gran Sasso, also other laboratories with underground accelerator projects are indicated: Canfranc/Spain \cite{CANFRANC}, Homestake/USA \cite{Robertson16-EPJWOC}, and Jinping/China \cite{Liu16-EPJWOC}. The underground depth is expressed in meters water equivalent, where the rock thickness is multiplied by the average rock density of typically 2.7\,g/cm$^3$.}\label{fig:muonflux}
\end{figure}

The LUNA laboratory is located in the Gran Sasso underground laboratory, shielded by 1400 meters of dolomite rock from cosmic ray induced effects. This Italian national laboratory is operated by the Italian National Institute for Nuclear Physics, INFN. The Gran Sasso site has several beneficial effects, the first of them is the complete suppression of the nucleonic component of cosmic rays.
The second, and most important, effect is the attenuation of the cosmic-ray muon flux by about six orders of magnitude when compared with the surface of the Earth (Fig.\ref{fig:muonflux}). Muons are highly penetrating elementary particles created in the upper atmosphere that leave a trace in most particle and $\gamma$-ray detectors and can in addition create unwanted spallation neutrons and radioactive nuclei in detector and shielding materials. It is therefore very important to note that at Gran Sasso both the direct and indirect effects of muons are negligible. The same is true for the other underground accelerator laboratories where accelerator projects are being prepared (Fig.\ref{fig:muonflux}), with the exception of the shallow-underground laboratory Felsenkeller in Germany, where the muon flux musts still be dealt with \cite{Szucs15-EPJA}.

The unwanted effects of muons have been studied in details in an intercomparison exercise, where one and the same high-purity Germanium (HPGe) detector was equipped with a BGO guard detector acting as a muon veto and tested first at the Earth's surface, then underground \cite{Szucs10-EPJA}. It was found that even though the muon veto improved the background level at the Earth's surface, still the background was 100 times lower at Gran Sasso. This finding \cite{Szucs10-EPJA} underlined the difficulty of removing muon effects only using technical means, a problem that is essentially absent at Gran Sasso due to the underground location.

Another source of background are the direct and indirect effects of long-lived radioisotopes in the laboratory and the experimental setup.
The most notable problem in this line are the $\gamma$-rays emitted by the natural $^{238}$U and $^{232}$Th decay chains and from $^{40}$K. These effects are present in any laboratory and do not depend on depth, but rather on the radiopurity of building and detector materials. They can be mitigated by a suitable passive shielding surrounding the target and the detectors, usually consisting of selected low-background lead and freshly refined electrolytic copper.  Impurities in the detector and target, on the other hand, must be removed by proper material selection \cite{Heusser95-ARNPS}.

One problem that is only apparent in well-shielded setups may be bremsstrahlung generated by $\beta$ emitters in the shielding, for example $^{210}$Bi (decay product of $^{210}$Pb) in lead bricks \cite{Caciolli09-EPJA}; this leads to limits on the permissible level of $^{210}$Pb contamination in lead shields. Another problem is radioactive radon gas, which emits a number of $\gamma$ rays when decaying. This can be dealt with by enclosing target and detector in an airtight anti-radon box that is flushed with radon-free gas, for example from the steam-off of liquid nitrogen cooling HPGe detectors.

\begin{figure}[h!]
\centerline{%
\includegraphics[width=0.8\textwidth]{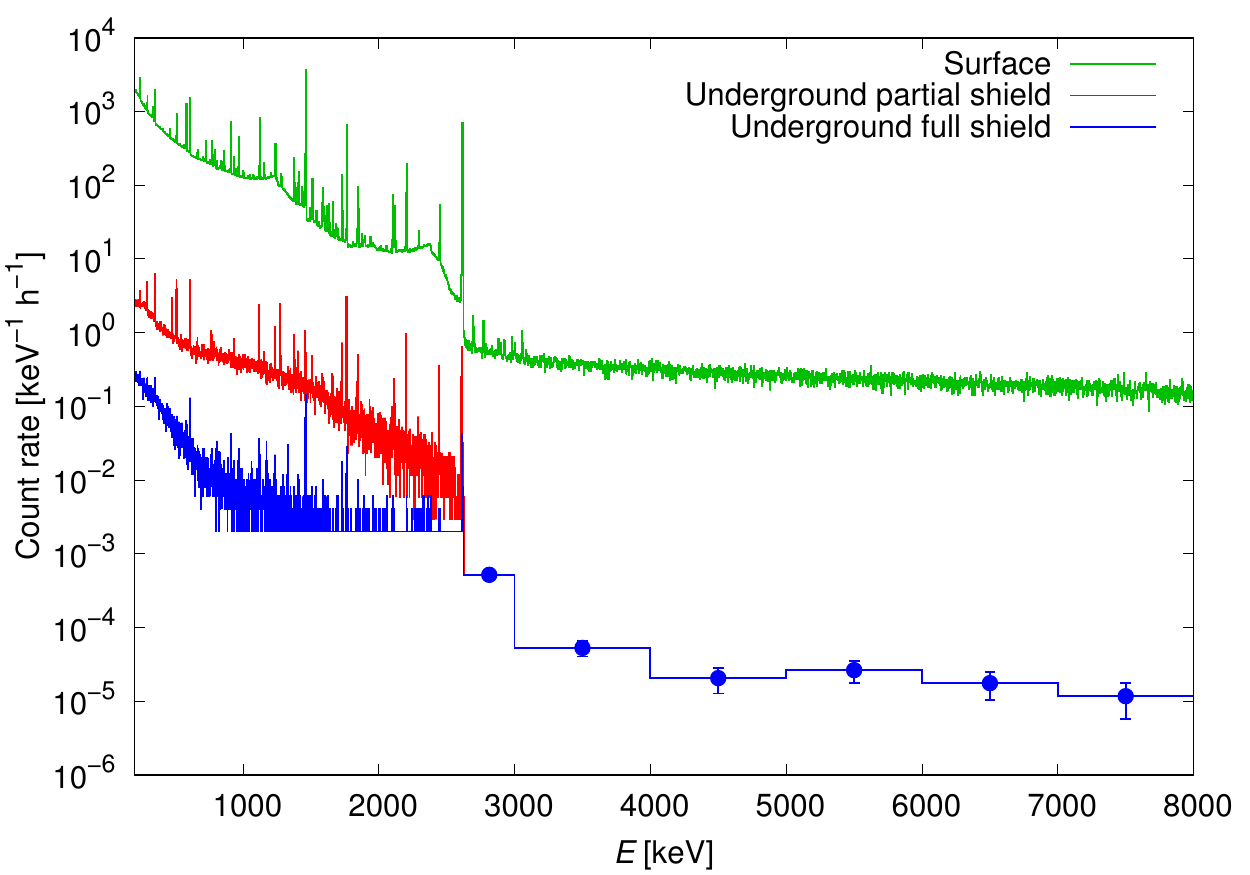}}
\caption{Laboratory background measured with the germanium detector: at the surface, underground at LNGS with the full shield (Cu+Lead+anti-Rn \cite{Caciolli09-EPJA}) and with
the partial shield (Cu+Pb and a second Ge detector inside the shield \cite{Cavanna14-EPJA}).
\label{fig:fondo_GePD}}
\end{figure}

The single $\gamma$-ray and bremsstrahlung effects are contained in a limited energy region,  $E_\gamma \leq$ 2.615\,MeV, which is the energy of the $\gamma$ from the decay
of $^{208}$Tl in the  $^{232}$Th chain. For the deep-underground setting of LUNA, a shielding of 25\,cm lead with low $^{210}$Pb content lined at the inside with 5\,cm electrolytic copper has been found to give excellent background capabilities (Fig.\ref{fig:fondo_GePD}) \cite{Caciolli09-EPJA}. For lesser depths or overground laboratories, the optimum shielding thickness is typically lower \cite{Heusser95-ARNPS}, due to the compromise between background removed by the shielding and additional muon-induced background created in the shield.

Deep underground at Gran Sasso, muon-induced neutrons do not pose a problem on the level of background needed for nuclear astrophysics. However, at these depths a second source of neutrons that was completely buried by cosmic-ray effects at the Earth's surface emerges: Neutrons from spontaneous fission of $^{238}$U and ($\alpha$,n) reactions in the rock. In particular the latter effect is problematic, with the $\alpha$ particles supplied by the natural $^{238}$U and $^{232}$Th decay chains and the reactions taking place on several target nuclides in the rock \cite{Wulandari04-APP}. These reactions lead to a residual flux of energetic neutrons that is of the order of 4$\times$10$^{-7}$ cm$^{-2}$s$^{-1}$ for $E_{\rm n}>$1\,MeV \cite{Arneodo99-NCA}, about a thousand times below the neutron flux at the surface of the Earth.

This tiny flux of fast neutrons is the main driver of the residual background at LUNA, with effects that can be seen up to 12\,MeV in $\gamma$-ray detectors. For experiments searching for rare events \cite[e.g.]{Aprile16-PRD}, the neutron background is usually mitigated using shields to moderate and capture neutrons. However, for the purpose of nuclear astrophysics, higher signal counting rates are measured, and therefore the remaining background at LUNA proved sufficient for the experiments hitherto undertaken.

The environmental background status at LUNA has been studied in several publications addressing various aspects, namely: Background in a large BGO $\gamma$-calorimeter \cite{Bemmerer05-EPJA}, in an ultra-low-background HPGe detector surrounded by 5\,cm copper and 25\,cm lead \cite{Caciolli09-EPJA}, in an actively muon-shielded HPGe detector compared between Earth's surface and LUNA \cite{Szucs10-EPJA}, and in a silicon charged-particle detector \cite{Bruno15-EPJA}. In each case, an unprecedented low environmental background was found.

\subsubsection{Ion-beam induced background at LUNA}

Contaminants  in the setup material and in targets are not only problematic if they are radioactive, but also if they are reactive to the incident ion beam during the experiments.

The $^1$H$^+$ and $^4$He$^+$ beams produced by the LUNA accelerator may give rise to $\gamma$-ray producing nuclear reactions and, in special cases, also to the production of neutrons. As a general rule, contaminants with a nuclear charge that is higher than the nucleus under study in the experiment usually do not pose a problem, simply due to the action of the Coulomb barrier which generally suppresses the cross section also of the contaminant reactions.

Nuclear reactions that have given rise to backgrounds in proton-beam experiments include $^{2}$H(p,$\gamma$)$^{3}$He, $^{7}$Li(p,$\gamma$)$^{8}$Be, $^{11}$B(p,$\gamma$)$^{12}$C (resonance at $E_p$ = 163\,keV), $^{13}$C(p,$\gamma$)$^{14}$N, $^{18}$O(p,$\gamma$)$^{19}$F (resonance at 151\,keV), $^{19}$F(p,$\alpha\gamma$)$^{16}$O (resonances at 224 and 340\,keV), and $^{23}$Na(p,$\gamma$)$^{24}$Mg. The energy threshold for neutron production by the (p,n) channel is not reached by the LUNA 400\,kV accelerator.

\begin{figure}[h!]
\centerline{%
\includegraphics[width=0.8\textwidth]{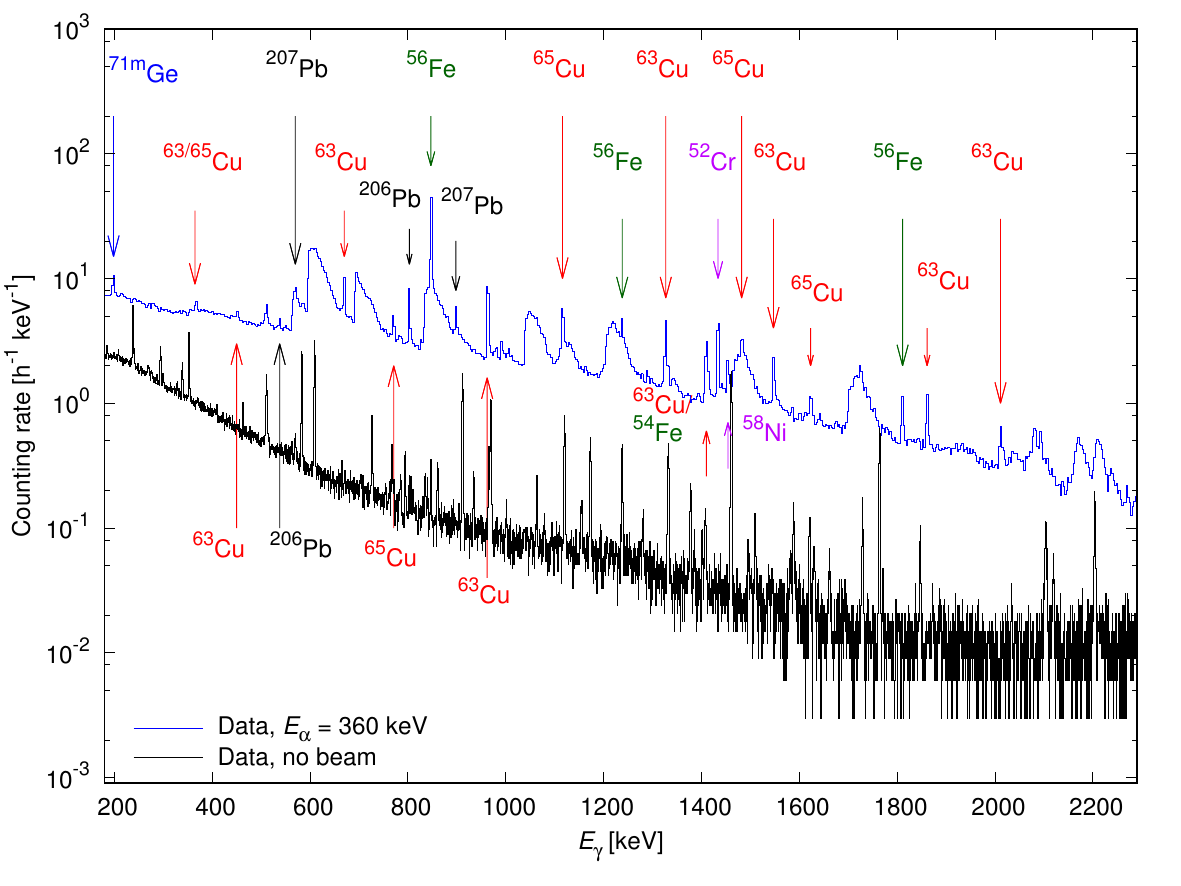}}
\caption{$^2$H($\alpha$,$\gamma$)$^6$Li spectrum taken with the germanium detector at E${_{\alpha}}$=360 keV  and target deuterium pressure of 0.3 mbar. The laboratory background, shown in black, has been subtracted. The detector set-up has not yet
been optimized to suppress the in-beam neutron production \cite{Anders13-EPJA}. The most important $\gamma$ lines due to (n,n',$\gamma$) and (n,$\gamma$) reactions on the different
materials are marked with arrows.}
\label{fig:Fig_neutrons}
\end{figure}

For the case of $\alpha$-beam experiments, no direct background emission has been observed at all at LUNA, neither for $\gamma$ rays nor for neutrons. However, an indirect two-step process proved very problematic in one case: The $\alpha$ beam created energetic deuterons by elastic scattering on a deuterium gas target. These energetic deuterons, in turn, produced neutrons by the efficient $^2$H(d,n)$^3$He process on other deuterium gas atoms \cite{Anders13-EPJA} (Fig.\ref{fig:Fig_neutrons}).

This background had to be carefully modeled and subtracted \cite{Anders13-EPJA,Anders14-PRL,Trezzi17-APP}.
If a contaminant reaction is observed, care must be taken to remove the relevant element from the setup, or even better exclude it already during the production process. Boron, carbon, oxygen, and fluorine are common elements that can be deposited due to several processes into the experimental setups. For example, fluorine is often found in target backings and can be deposited by accident, and the same may happen in the case of boron. Careful cleaning processes have been developed to reduce the contribution of these two elements \cite{Caciolli12-EPJA}.
Carbon, deuterium (hydrogen), and oxygen are usually found in the beam line as residual gases and they can be transported by the beam on the target surfaces. A suitable cold trap or at least highly efficient vacuum pumps may be used to reduce their amount.

\subsection{Accelerators}\label{subsec:accelerators}

The main characteristics of the ion accelerators used at LUNA are determined by the experimental needs: At least a factor of 10 dynamic range in acceleration potential is needed in order to cover a large part of the excitation function of a nuclear reaction under study. The acceleration potential must be stable to $\leq$1 keV over many hours in order to avoid changes in the yield due to the Coulomb barrier, and to $\leq$0.1 keV over one hour, in order to reliably perform energy scans of targets. Also, the energetic spread of the beam must be on a similar level, again to enable target studies by scanning the beam energy in small, sometimes sub-keV steps. The ion source must be able to run stably overnight without human intervention.

Two ion accelerators have hitherto been used at LUNA, and a third one will be installed soon. All three are of the electrostatic type, and the high voltage is supplied by Cockroft-Walton generators.

The first accelerator, the LUNA 50 kV machine, was operated from 1991 to 2001 \cite{Greife94-NIMA}. It included a duoplasmon ion source for intense hydrogen and helium beams. It is a hot cathode plasma source where an intense axial magnetic field is present in the discharge region providing a well-focused ion beam and consequently high ion current with low energy spread.

The main problem of this source is the relatively short lifetime, which made the experiment-maintenance intensive. Due to the low electric field used, the LUNA 50 kV accelerator worked in air.

In order to solve this issue, for the LUNA 400 kV accelerator (in operation since 2001 \cite{Formicola03-NIMA}), a radio-frequency ion source is used. It consists of a glass tube containing hydrogen or helium gas. A radio frequency field is applied, ionizing the gas inside the tube. The ions are then extracted from the ion source and accelerated. The whole accelerator is enclosed in a pressure tank filled with 20\,bar of a gas mixture consisting of nitrogen and carbon dioxide.

The third LUNA accelerator, LUNA MV, will be installed in 2018 \cite{Guglielmetti14-PDU,Broggini17-EPJWC,Formicola2017-JPSCP}. Detailed information on the accelerator can be found in \cite{LUNA15-AR,Aliotta16-EPJA}. Briefly, the acceleration potential of 3.5\,MV is an order of magnitude larger than that of the LUNA 400 machine. The new accelerator will be able to provide not only intense $^1$H$^+$ and $^4$He$^+$ beams, but also  carbon beams (single and double-charged).
After the accelerator, the ion beam passes through an electromagnet used as a beam analyzer. The  LUNA MV accelerator, as the
LUNA 400 kV one,
is equipped with two different beam lines, only one of them fed at a time.

The accelerator room will have thick concrete walls and ceiling working as neutron shielding. The maximum
neutron flux just outside the shielding, averaged over the entire external surface, will be significantly smaller than the natural neutron background at LNGS, with a similar energy spectrum.
\begin{table}[h]
\begin{center}
    \begin{tabular}{|l | l r | c | c | c | c | c|}
    \hline
     \textbf{} & \textbf{Ion} & {\bf Current} & \textbf{Terminal Voltage} & \textbf{Stability} \\
     \textbf{} & & \textbf{($\mu$A)} & \textbf{(kV)} & \textbf{(eV/h)} \\
    \hline
    \hline
     					 & $^1$H$^{+}$ & 1000 &  & \\
    \textbf{LUNA-50}	 & $^3$He$^{+}$ & 500 & 3-50 & $<$ 5 \\
    					 & $^4$He$^{+}$ & 500 &  & \\
    \hline
    \hline
    					 & $^1$H$^{+}$ & 1000 &  &  \\
    \textbf{LUNA-400} 	 & $^3$He$^{+}$ & 500 & 50-400 & $<$ 5 \\
    					 & $^4$He$^{+}$ & 500 &  & \\
    \hline
    \end{tabular}
\caption{The LUNA 50 kV and LUNA 400 kV accelerator parameters. The ion currents listed are the maximum available at the end of the accelerator tube.}\label{tab:accelerators}
\end{center}
\end{table}

\subsection{Target assemblies}\label{subsec:targets}

For a precision cross section measurement, as is usually the case at LUNA, it is imperative to know the target in details. At the low energies studied here, the target usually serves also as the beam stop. Therefore a second important input for cross section determinations, the measurement of the ion beam intensity, has to be considered together with the target design.

For LUNA experiments, usually a target with larger than the ion beam diameter is selected, so to have the beam fully contained inside the target. Then the target thickness, stoichiometric and isotopic composition should be precisely known. The intensive ion beam from the 400 kV accelerator may deposit up to 200\,W in the target, so the target stability under bombardment must be maintained and monitored.

In addition, a careful treatment for removing and monitoring possible contaminants in the targets is necessary, especially in long term measurements such as those of LUNA, where contaminants could not only create problematic structures in the acquired spectra, but also modify the target structure introducing passive layers.

The following considerations are limited to the present LUNA 400 accelerator, because the relevant points for target systems are not much different for the cases of the previous LUNA 50 and the future LUNA MV accelerators. LUNA 400 is equipped with two beam lines:  The first one includes a windowless gas target, and the second one a solid target assembly.

\subsubsection{Gas Targets}\label{sec:gastarget}

A windowless gas target has been the target of choice for a number of reaction studies at LUNA. First and foremost, gas targets offers stability over the long running times, up to several weeks per data point, needed. Secondly, for the nuclear reactions under study at LUNA, solid targets never consist of one element alone: Either they are chemical compounds of several elements, or they are one element (e.g. a noble gas) implanted into a backing consisting of other elements. In both cases, a significant part of the energy loss is due to the "passive" element. At very low energies, when a few keV ion beam energy loss may correspond to a factor of 10 drop in cross section, this limits the luminosity and is thus undesirable.

\begin{figure}[htb]
\centerline{%
\includegraphics[scale=0.5]{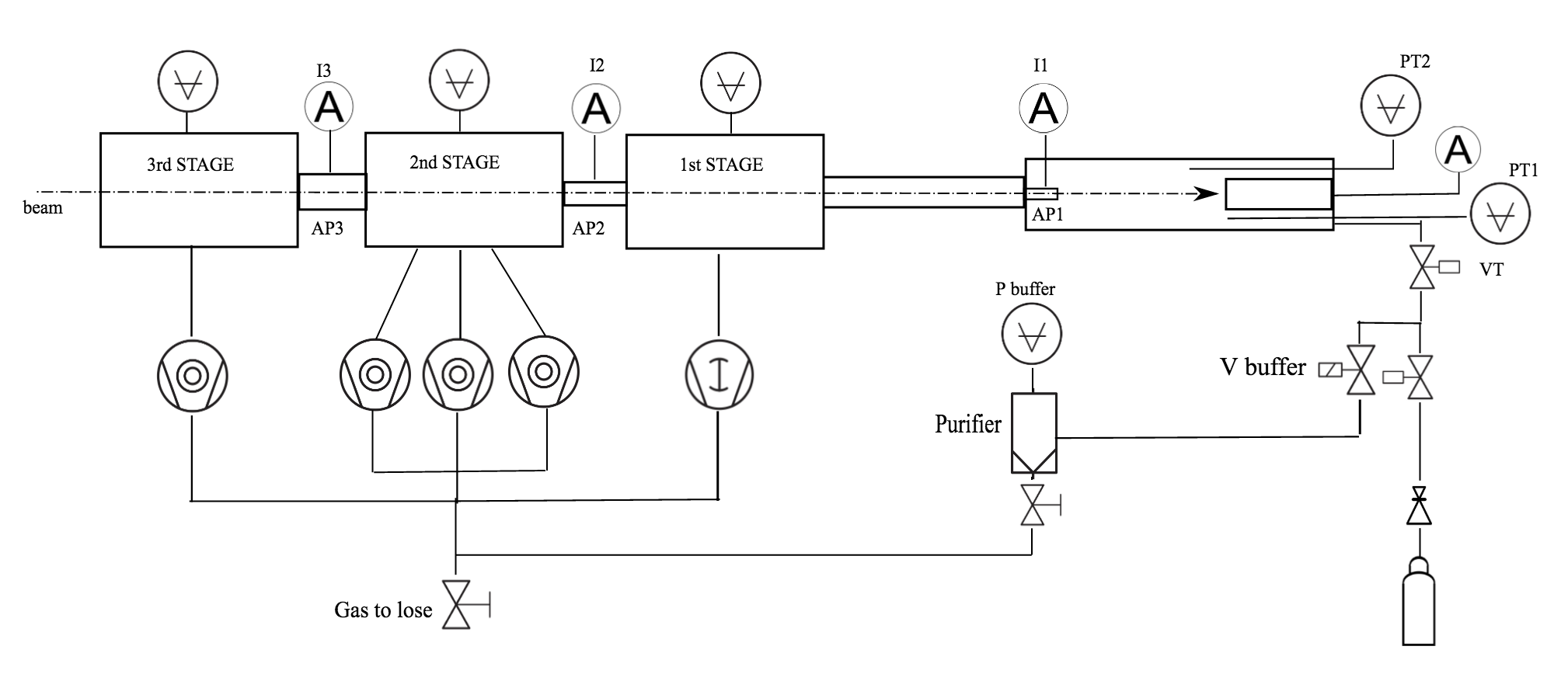}}
\caption{Simplified scheme of the LUNA gas target. More details can be found in \cite{Anders13-EPJA,Cavanna14-EPJA}.}\label{fig:gas_target}
\end{figure}

The heart of the LUNA windowless gas target system is a 10-40\,cm long scattering chamber. The selected gas is pumped out of the scattering chamber through a several cm long, narrow tube which also serves as entry collimator to allow the ion beam to enter the gas target. The size of this collimator is a careful compromise between beam optics (favoring a wide, short tube for easier focusing) and gas flow impedance (favoring a narrow, long tube for lower gas consumption). The gas is replenished through another entry to the scattering chamber, either from a gas bottle or from a recirculation system, depending on the experiment, with a feedback-controlled valve ($V_T$ in Fig.\ref{fig:gas_target}) ensuring that the pressure in the target chamber remains constant.

A differential pumping system including Roots pumps, very large turbomolecular pumps, and oil-free forepumps ensures that even though the scattering cell is not physically separated from the accelerator, the vacuum in the accelerator remains at the 10$^{-7}$ mbar level even for target gas pressures up to 10\,mbar. The exhaust from the forepumps is either discarded or purified in a chemical getter and recirculated to the target, depending on the experiment and gas used.

A schematic picture of a typical recirculating gas target setup installed at the LUNA facility is shown in Fig.\ref{fig:gas_target}.

\begin{figure}[htb]
\centerline{%
\includegraphics[scale=1.1]{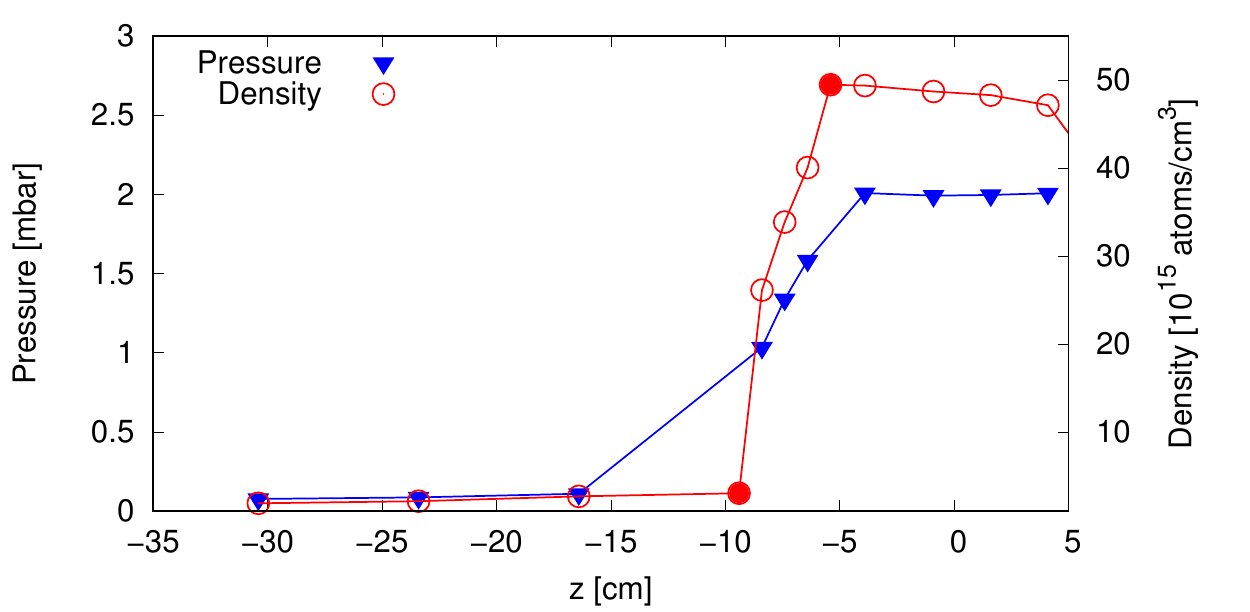}}
\caption{The pressure profile (triangles) measured with natural neon at 2 mbar in the gas target together with the reconstructed density profile (empty circles). The value of the density at the beginning and at the end of the collimator is obtained from extrapolation (closed circles). The center of the scattering chamber, 10 cm long, is at z=0.}\label{fig:gas_profiles}
\end{figure}

For the proper characterization of a windowless gas target, the density profile along the beam path  must be known. The density, in turn, depends on the temperature and pressure which have to be measured along the whole beam path. This is usually done using a mock-up scattering chamber equipped with measurement ports for capacitive pressure gauges and thermoresistors. For the final experiment, the mock-up chamber is replaced by a chamber with equal dimensions but without the measurement ports, in order to make space for the $\gamma$ detector and its shielding. A typical density profile, obtained by combining the pressure and temperature profiles, is shown in Fig.\ref{fig:gas_profiles}.

Two effects specific to gas targets must then be corrected for. First, the beam loses energy, and thus heats the target gas as it passes through it. This effect can quickly reach corrections of several percent and must thus be measured. For the case of reactions with narrow, strong resonances, this can be done by the resonance scan technique, where the thinning of the gas is monitored by the effective position of a sharp resonance in the target chamber. At LUNA, resonances in $^{14}$N(p,$\gamma$)$^{15}$O \cite{Bemmerer06-NPA} and $^{21}$Ne(p,$\gamma$)$^{22}$Na \cite{Cavanna15-PRL} have been used for this purpose.
A second option is the elastic scattering yield, which has been used to monitor a helium gas target at LUNA \cite{Marta06-NIMA}.

The second effect specific to a gas target is the impossibility of an electric beam current measurement. The charge state of the ion beam is gradually altered from its original singly-positive state as it passes the various pumping stages and the gas target itself. In addition, a high flux of electrons due to ionization of gas molecules is observed on all current meters. Therefore, the beam intensity is measured by a beam calorimeter with constant temperature gradient. The heat load is supplied both by the ion beam hitting the beam calorimeter and by regulated heating resistors. A feedback loop monitoring the temperatures controls the resistors so that the temperatures are always constant, thus guaranteeing that the sum of beam power and resistor power is also constant.

The beam intensity, $I$, is then calculated as
\begin{equation}\label{eq:calorimeter}
I = \frac{W_{0} - W_{\rm run}}{E_p - \Delta E_p}
\end{equation}
where $W_0$ and $W_{\rm run}$ are the resistor heating powers measured without and with beam, respectively, and $\Delta E_p$ is the energy loss the ion beam of energy $E_p$ experiences before hitting the calorimeter surface.

Contaminants in the gas are usually negligible, but for selected cases (like in-leaking air) they can be monitored using nuclear reactions. There may be beam induced background  from contaminations on the surfaces of the scattering chamber that are hit by the beam, which have to be mitigated by cleaning the surfaces and avoiding hydrocarbons in the vacuum.

The LUNA gas target systems used for the various nuclear reactions studied with a gas target have been described in a number of original publications including more details: $^{3}$He($^{3}$He,2p)$^{4}$He \cite{Junker98-PRC}, $^2$H(p,$\gamma$)$^3$He \cite{Casella02-NIMA}, $^{14,15}$N(p,$\gamma$)$^{15,16}$O \cite{Bemmerer06-NPA}, $^{3}$He($\alpha$,$\gamma$)$^{7}$Be \cite{Caciolli09-EPJA}, $^{2}$H($\alpha$,$\gamma$)$^{6}$Li \cite{Anders13-EPJA}, and $^{22}$Ne(p,$\gamma$)$^{23}$Na \cite{Cavanna14-EPJA}.

Gas jet targets may in the future provide an alternative to the windowless gas targets currently in use at LUNA; however they pose many challenges especially in the determination of the target thickness and in the definition of the target edges \cite{Chipps14-NIMA}.

\subsubsection{Solid Targets}\label{sec:solidtarget}

Solid targets permit more compact experimental setups than gas targets, allowing to install the detector in a  closer geometry. They can usually be approximated as a point-like source, which is convenient when studying the angular distribution of emitted $\gamma$ rays. In addition, solid targets can be used for all the cases where gas targets are impractical. This is notably also possible for gaseous elements, by using a chemical compound or by implanting them into a metal matrix.

The target usually presents itself in the form of a thin, self-supporting metallic disk with the target material itself either deposited on the metal backing or implanted into it. The disk form is necessary, because the target is used to stop the ion beam and must thus be watercooled to remove the significant heat load by the ion beam. This form also allows for the convenient change from one target to another, as is necessary when the ion beam has degraded a given target.

\begin{figure}[htb]
\centerline{%
\includegraphics[scale=1]{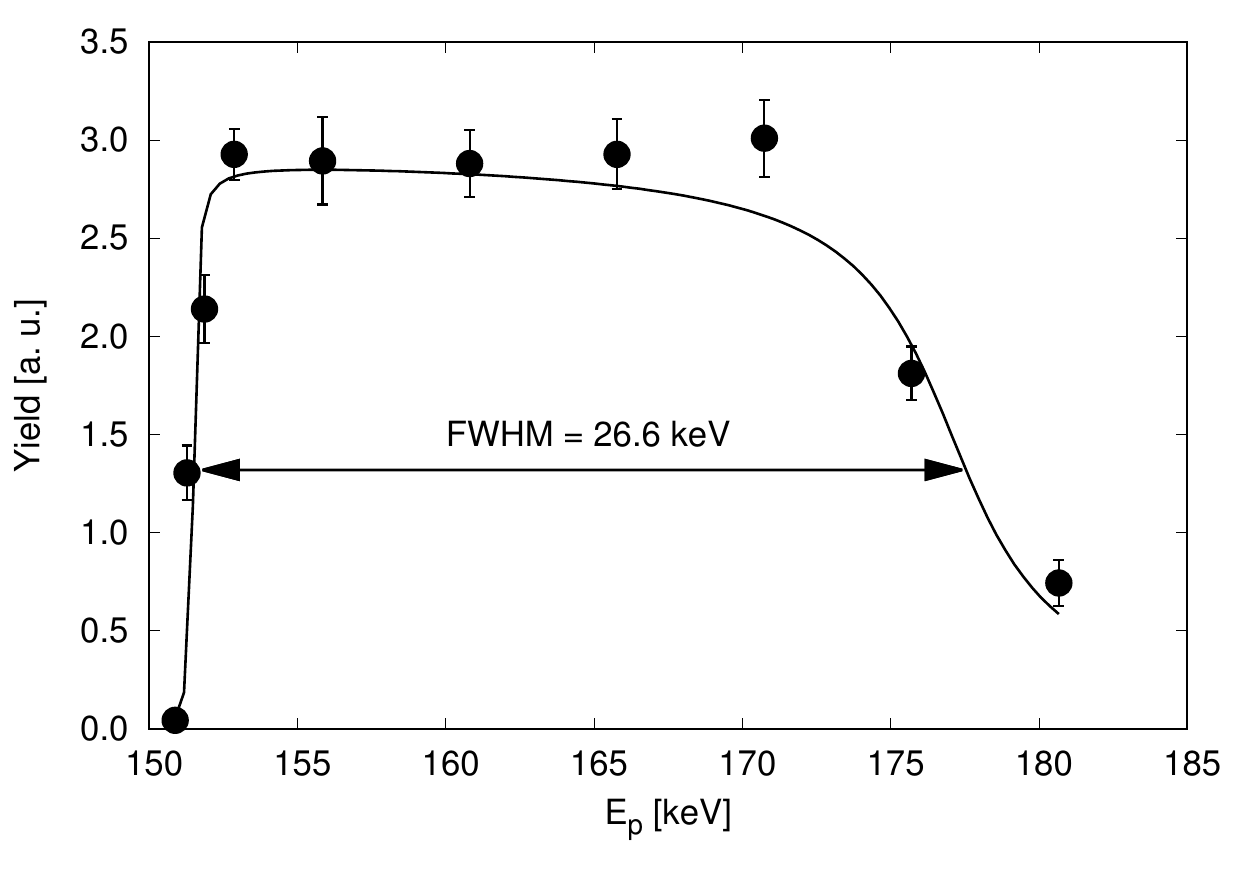}}
\caption{Scan of the 151 keV resonance of the $^{18}$O(p,$\gamma$)$^{19}$F reaction performed on a Ta$_2$O$_5$ target. A fit used to determine the target thickness is also shown.}\label{fig:thickness}
\end{figure}

Before hitting the target, the ion beam has to pass
a cold trap made of a $\sim$1\,m long copper tube cooled with liquid nitrogen that reaches up to a few mm distance from the target surface. This way, contaminants from the residual gas freeze out on the copper, thus suppressing the build-up of impurities during irradiations that may run continuously for several weeks \cite{DiLeva14-PRC,Caciolli11-AA}. In addition, the cold trap is negatively charged to force secondary electrons emitted on the target back on the target surface. The ion beam current can then be measured electrically, using the Faraday cup given by the target and parts of the target chamber.

At LUNA, the backing is directly watercooled by deionized water, ensuring that the electrical current lost through the cooling water can be neglected. A favored backing material is tantalum, which is an efficient heat conductor to connect the target via the backing to the cooling water.

Targets are produced using different techniques.  The most common ones are evaporation, reactive sputtering, oxidation, and implantation.
For a particular experiment, the selection of the technique is  based on the target nucleus under study and its chemical behaviour.

Different from the gas target, the characteristics of a solid target, namely exact stoichiometric ratio, thickness, and concentration of impurities, can not be exactly determined from the production process but must be measured after production.
In case  additional contaminants are found on the backings, appropriate cleaning solutions must be applied \cite{Caciolli12-EPJA}.

For the determination of the stoichiometric ratio, in many cases it is preferable to use a method of ion beam analysis \cite{Nastasi15-Book}. The most precise results are usually obtained by nuclear reaction analysis or its subtype, nuclear resonant reaction analysis. Examples used at LUNA include notably the nitrogen concentration based on the resonant $^{14}$N(p,$\gamma$)$^{15}$O reaction. In case of an energetically thin (proton width $\Gamma_p \ll \Delta E_p$, with $\Delta E_p$ the energy loss over the total target thickness) target, so-called target scans are applied (Fig.\ref{fig:thickness}). They allow to measure directly the energetic thickness $\Delta E_p$ of the target.

In cases where in-situ methods cannot be used, a variety of ion beam analysis methods have been applied, both on virgin targets and on targets after irradiation at LUNA. To this end, the targets were dismounted and transported to other facilities for analysis.
Techniques used include Rutherford backscattering (RBS),  elastic recoil detection analysis (ERDA) including its heavy-ion variant that can disentangle the target contribution also from a heavy backing such as tantalum, and  secondary ion mass spectroscopy (SIMS) including its variant SNMS (secondary neutral mass spectrometry). A detailed description of several techniques applied and their application on  Ta$_2$O$_5$ targets isotopically enriched in $^{17}$O can be found in \cite{Caciolli12-EPJA}.

\subsection{Detectors}\label{subsec:detectors}

For the sake of the discussion, first $\gamma$-ray detectors will be reviewed. The detectors used at LUNA can be classified into spectroscopy-oriented detectors, where attention is paid to angular and energy resolution, and calorimetry-oriented detectors, where the highest achievable $\gamma$-ray detection efficiency is used so that branching and angular effects are diluted. For the first category, high-purity germanium (HPGe) detectors are used, usually monolithic but in some cases segmented. For the second category, a 70\,kg bismuth germanate (BGO) borehole detector of 7\,cm outer thickness is available. This detector forms a cylindrical 4$\pi$ detector  surrounding the target position \cite{Casella02-NIMA} and reach an efficiency close to 70\% even for $E_\gamma$ = 10\,MeV, in addition to its calorimetric capabilities.

In cases where it is possible to directly take data at Gamow peak energies, the total  cross section is the only parameter needed to compute the thermonuclear reaction rate. In that case, for negligible ion beam background, the calorimetric BGO detector is the best choice, and it has been used for the $^{14}$N(p,$\gamma$)$^{15}$O \cite{Lemut06-PLB,Bemmerer06-NPA}, $^{15}$N(p,$\gamma$)$^{16}$O \cite{Bemmerer09-JPG,Caciolli11-AA}, and $^{25}$Mg(p,$\gamma$)$^{26}$Al \cite{Strieder12-PLB} cases.

In cases where extrapolations are still needed, it is necessary to evaluate each "transition", i.e. the capture to each excited state in the compound nucleus, separately. This necessitates a "spectroscopic" HPGe detector, as it has been used for the $^{14}$N(p,$\gamma$)$^{15}$O \cite{Formicola04-PLB,Imbriani05-EPJA,Marta08-PRC,Marta11-PRC}, $^{3}$He($\alpha$,$\gamma$)$^{7}$Be \cite{Confortola07-PRC}, $^{25}$Mg(p,$\gamma$)$^{26}$Al \cite{Limata10-PRC}, $^{15}$N(p,$\gamma$)$^{16}$O \cite{LeBlanc10-PRC}, $^{17}$O(p,$\gamma$)$^{18}$F \cite{Scott12-PRL,DiLeva14-PRC}, and $^{22}$Ne(p,$\gamma$)$^{23}$Na \cite{Cavanna15-PRL,Depalo16-PRC} cases.

For both the HPGe and BGO cases, the efficiency is usually determined by a Monte Carlo simulation, validated with experimental measurements with calibrated point-like $\gamma$-activity standards such as $^{137}$Cs, $^{60}$Co, $^{22}$Na, and $^{88}$Y. Using the two-line method with the 278\,keV resonance in the $^{14}$N(p,$\gamma$)$^{15}$O reaction which includes three 1:1 cascades with isotropic $\gamma$-ray emission through the levels at 5.182, 6.172, and 6.792\,MeV, the efficiency data can be extended up to 7\,MeV. If the Monte Carlo matches both the low-energy data from the activity standards and the high-energy data from the two-line method, it is generally assumed to apply also at lower and higher energies, where the underlying detector physics is still comparable.

A special case is the activation method, which has been used for the $^{3}$He($\alpha$,$\gamma$)$^{7}$Be \cite{Bemmerer06-PRL,Confortola07-PRC,Gyurky07-PRC} and $^{17}$O(p,$\gamma$)$^{18}$F \cite{Scott12-PRL,DiLeva14-PRC} cases. In the first case, the kinematically forward focused $^7$Be was collected on a copper catcher in the $^3$He gas target; in the second case, the $^{18}$F was stopped inside the solid target itself. The sample was then dismounted from LUNA and brought to the Gran Sasso low-activity counting setup \cite{Laubenstein04-Apradiso}. There, activity measurements down to 90 mBq were performed with $^{18}$F (half-life: 110 minutes) and sub-mBq activity measurements were possible for $^7$Be (half-life: 53.22 days).

A second special case are particle detectors. LUNA-50 was first developed in order to enable ultra-low background in $\Delta E$-$E$ silicon telescopes and single silicon detectors used for the $^{3}$He($^{3}$He,2p)$^{4}$He reaction \cite{Junker98-PRC,Bonetti99-PRL}. Single silicon detectors were also used for the study of the $^{17}$O(p,$\alpha$)$^{14}$N  \cite{Bruno15-EPJA,Bruno16-PRL} and $^{18}$O(p,$\alpha$)$^{15}$N reactions.

\section{Astrophysical sites for nucleosynthesis}\label{sec:sites}

The experimental technique employed at LUNA, direct experiments with high-intensity beams of stable ions, is ideally suited for two scenarios: First, the study of hydrostatic burning that takes place during long, quiescent periods in the history of a star, at temperatures of 0.01 - 0.1 GK. Second, explosive burning at relatively moderate temperatures up to 1 GK, in sites such as the Big Bang, astrophysical novae, or some situations found previous to or during thermonuclear supernovae.

The classical examples that are usually cited for hydrostatic burning processes are two processes of hydrogen burning: The proton-proton chain and the carbon-nitrogen-oxygen (CNO) cycle, both of which take place in young or low-mass stars in a unique, central burning zone. These processes have been studied in details at LUNA; both the astrophysics and the experimental situation are discussed in details in section \ref{sec:solar}.

In addition to this classical scenario, the CNO cycle can also take place in the hydrogen-burning shell of a more complicated star which is characterized by a layered structure, like that of an onion. Such a structure evolves for stars that are as massive as our Sun, or even more massive, after the completion of core hydrogen burning.
However, outside the center there is still enough hydrogen to feed catalytic hydrogen burning processes in a so-called hydrogen burning shell. These processes can also include higher processes of hydrogen burning, such as the second and third CNO cycles, neon-sodium, and magnesium-aluminum cycles. A large number of the nuclear reactions driving hydrogen shell burning have been studied at LUNA, and they are reviewed in section \ref{sec:AGB}.

The hydrogen-depleted core then contracts and heats up, eventually igniting core helium burning with its higher characteristic temperatures.
The nuclear reactions of the helium-burning stage were able to bridge for the first time  the mass gap at atomic mass $A$ = 8, where Big Bang nucleosynthesis ceased. It was then that much of the carbon and oxygen,  essential ingredients for human life, was created.

In addition to the products of the hydrogen and helium burning zones, when taken alone,  some exchange of nucleosynthetic material between these zones may occur in onion-like stars. This phenomenon gives rise to a rich nucleosynthetic scenario, believed to be responsible for the creation of a great number of chemical elements, up to bismuth, by means of the so-called slow neutron capture process or astrophysical s-process \cite{Burbidge57-RMP,Cameron57-PASP,Kaeppeler11-RMP}. This process takes its name from the fact that the stellar scenario is such that neutron capture proceeds slowly compared to the radioactive $\beta^-$ decay of the product nuclei. This condition is typically fulfilled for a neutron density of 10$^{8}$-10$^{10}$ cm$^{-3}$. This sets it apart from the astrophysical r-process (r for rapid), where higher neutron densities ($\geq$10$^{20}$ cm$^{-3}$) cause neutron capture to proceed more rapidly than the radioactive $\beta^-$ decay of the capture products, leading to a nucleosynthetic path
which runs basically through beta-unstable nuclei close to the
neutron-drip line.
The r-process is expected to occur mainly in the extreme conditions typical of neutron star mergers.

\begin{figure}
\centering
\includegraphics[width=0.9\textwidth]{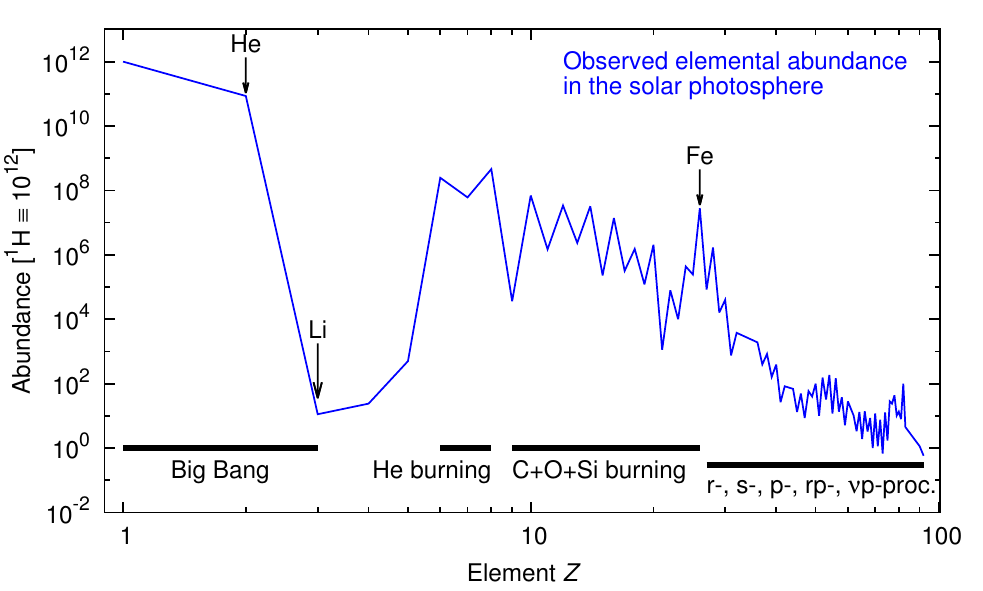}
\caption{The abundance of the chemical elements in the photosphere of the Sun. The astrophysical processes mainly responsible for the observed abundances are listed.}\label{fig:abundance}
\end{figure}

For the second class of astrophysical scenarios addressed at LUNA, i.e. explosive burning at relatively low temperatures, the textbook example is Big-Bang nucleosynthesis \cite{PDG2016}. It produces only three chemical elements: hydrogen ($^1$H and $^2$H), helium ($^3$He and $^4$He), and lithium ($^7$Li and traces of $^6$Li). There are no stable nuclei of atomic mass 8, and this so-called mass gap prevents Big-Bang nucleosynthesis from producing significant amounts of nuclei beyond $^7$Li. Big-Bang hydrogen and helium are the fuel for the quiescent hydrogen and helium burning processes discussed above. The main Big-Bang nuclear reactions have all been studied at LUNA, see section \ref{sec:BBN}.

So-called white dwarf stars (composed of carbon and oxygen, all ashes of helium burning) give rise to additional processes of explosive burning when they accrete material from a companion star, or when they merge with another white dwarf. Their signature on the night sky is usually of much shorter duration than a human lifespan and called an astrophysical nova (latin for new) or thermonuclear supernova. During explosive hydrogen burning in a nova, the temperatures may reach 0.4 GK, and in addition to the higher burning processes discussed above in the context of a layered, onion-like star, also proton capture on radioactive nuclei becomes possible in the hot-CNO and rp-process (rapid proton capture) scenarios. While the former case has been extensively studied at LUNA (section \ref{sec:AGB}), the two latter cases cannot be studied there because they involve radioactive ions.

In a thermonuclear type Ia supernova, explosive carbon burning drives the initial energetics and nucleosynthesis.
Type Ia supernovae are violent events originating from mass accretion onto a white dwarf in close binary systems. The mass increase is accompanied by a temperature increase which continues until carbon in the white dwarf's core ignites, sending a carbon deflagration wave through the core which leads to the supernova explosion.
This process is initially controlled by two main reactions involving $^{12}$C, the $^{12}$C($^{12}$C,p)$^{23}$Na and $^{12}$C($^{12}$C,$\alpha$)$^{20}$Ne reactions. These reactions play a role in quiescent burning, as well: In stars that are massive enough so that after the exhaustion of, first, core hydrogen burning, and, second, core helium burning, there is enough gravitational energy to heat the carbon-oxygen core enough to ignite carbon burning, there may be core carbon burning. Even heavier stars may become more stratified still, leading to a carbon-burning shell. These carbon-burning processes are highly uncertain and a mainstay of the program of the future LUNA MV accelerator. They are reviewed in section \ref{sec:c12c12}.

Seen from the perspective of the relative contribution to the final elemental abundances observed in the solar system (Fig.\ref{fig:abundance}), LUNA measurements address most of the quiescent and explosive processes leading up the iron abundance peaks. In addition, the planned studies on the neutron sources for the astrophysical s-process affect about half of the elements created beyond the iron peak. The LUNA experiment, in its deep underground location, therefore finds itself at the center of the study of the creation of the chemical elements in the Universe.
\section{Big Bang Nucleosynthesis}\label{sec:BBN}

\begin{figure}[h!]
    \begin{center}
  \includegraphics[scale=1.4]{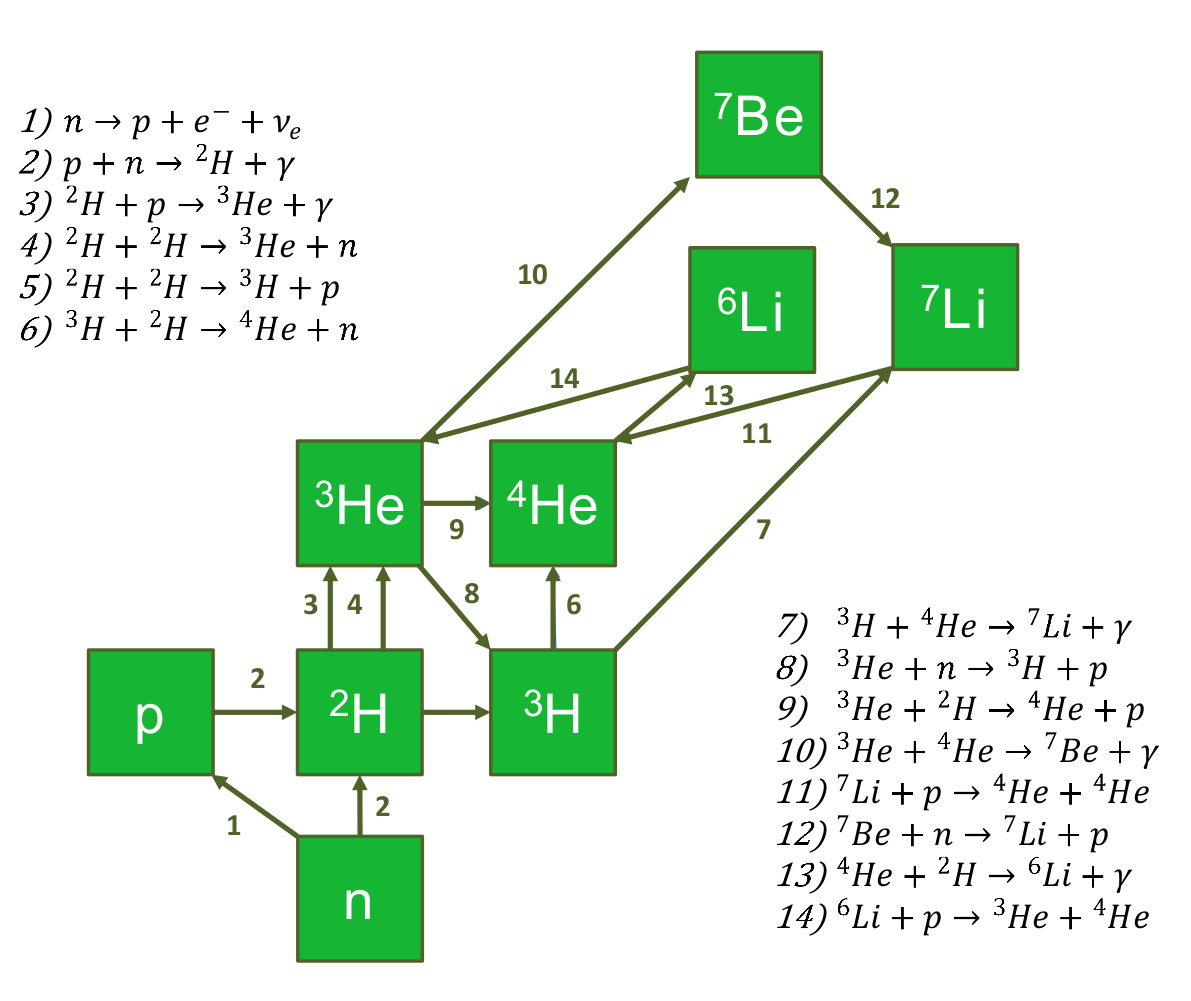}
  \caption{Nuclear reactions involved during the BBN era.}
\label{BBN}
    \end{center}
\end{figure}

The Lambda Cold Dark Matter ($\Lambda$CDM) model provides the simplest description of the Universe evolution and it is nowadays considered the standard cosmological model. It is  based on three experimental evidences: the cosmic expansion, the Cosmic Microwave Background radiation (CMB) and the primordial isotope abundances.
In particular, Big Bang Nucleosynthesis, BBN, is depicted as starting three minutes and half after the Big Bang, when the temperature of the Universe was low enough to produce deuterons through the $^1$H(n,$\gamma$)$^2$H reaction avoiding its destruction through the photo-dissociation process $^2$H($\gamma$,n)$^1$H. In about 20 minutes light nuclei such as $^2$H,$^3$H,$^3$He,$^4$He,$^6$Li,$^7$Li and $^7$Be have been created, following the reaction network reported in Fig.\ref{BBN}. Temperature and density moved in the
range T $\sim$ 0.9-0.4 GK and $\rho$ $\sim$ 2$\cdot$10$^{-5}$-2$\cdot$10$^{-6}$ g$\cdot$cm$^{-3}$, respectively.

The primordial abundances depend on the cross sections and on the barion to photon ratio, $\eta$, which is the only free parameter for BBN in the standard cosmological model.
As a consequence, by using the $\eta$ extracted from the CMB  measurements of the PLANCK satellite \cite{Planck15} and the existing values of the cross sections it is possible to calculate primordial abundances.
Predictions are then compared to the abundances obtained from the high resolution electromagnetic spectra of primordial objects and/or of astrophysical environments.
Table \ref{tab:abundances} gives the comparison between  predictions and measurements.

$^4$He is the only isotope whose abundance strongly depends on the expansion rate of the Universe.
The primordial abundances of the other isotopes depend only on the BBN reaction network (for a given $\eta$), i.e. on the nuclear cross sections involved. Unfortunately these isotopes appear only as small quantities, usually very difficult to be spectroscopically measured.
Deuterium is the only isotope whose primordial abundance has been measured with high accuracy. Unfortunately, in this case the prediction is affected by the relatively large uncertainty due to the $^2$H(p,$\gamma$)$^3$He cross section.
\begin{table}[h!]
\begin{center}
   \begin{tabular}{| c | c | c |}
   \hline
   \textbf{Isotope} & \textbf{BBN Calculations}  & \textbf{Observations} \\
   \hline
   \hline
   Y$_p < 1$ & 0.24709 $\pm$ 0.00025 \cite{Cyburt16-RMP}& 0.2449 $\pm$ 0.0040 \cite{Aver15-JCAP} \\
   $[$D/H$]$ & (2.65 $\pm$ 0.07)$\times$10$^{-5}$ \cite{DiValentino14-PRD} & (2.53 $\pm$ 0.04) $\times$ 10$^{-5}$ \cite{Cooke14-ApJ} \\
   $[^3$He/H$]$ & (1.0039 $\pm$ 0.0090) $\times$ 10$^{-5}$ \cite{Cyburt16-RMP} & (1.1 $\pm$ 0.2) $\times$ 10$^{-5}$ \cite{Bania02-Nature} \\
   $[^7$Li/H$]$ & (4.68 $\pm$ 0.67) $\times$ 10$^{-10}$ \cite{Cyburt16-RMP} & (1.58 $^{+0.35}_{-0.28}$) $\times$ 10$^{-10}$ \cite{Sbordone10-AA} \\
   \hline
   \end{tabular}
\caption{BBN calculated and astronomical observed abundances of primordial isotopes. Here, the $^4$He abundance is given in terms of the baryon mass fraction Y$_p$ i.e. the ratio between helium and baryon density. The abundance of the other isotopes is expressed by ratios.}\label{tab:abundances}
\end{center}
\end{table}

\subsection{Deuterium burning}\label{sec:deuterium}

\begin{figure}[h]
    \begin{center}
  \includegraphics[scale=0.7,angle = -90]{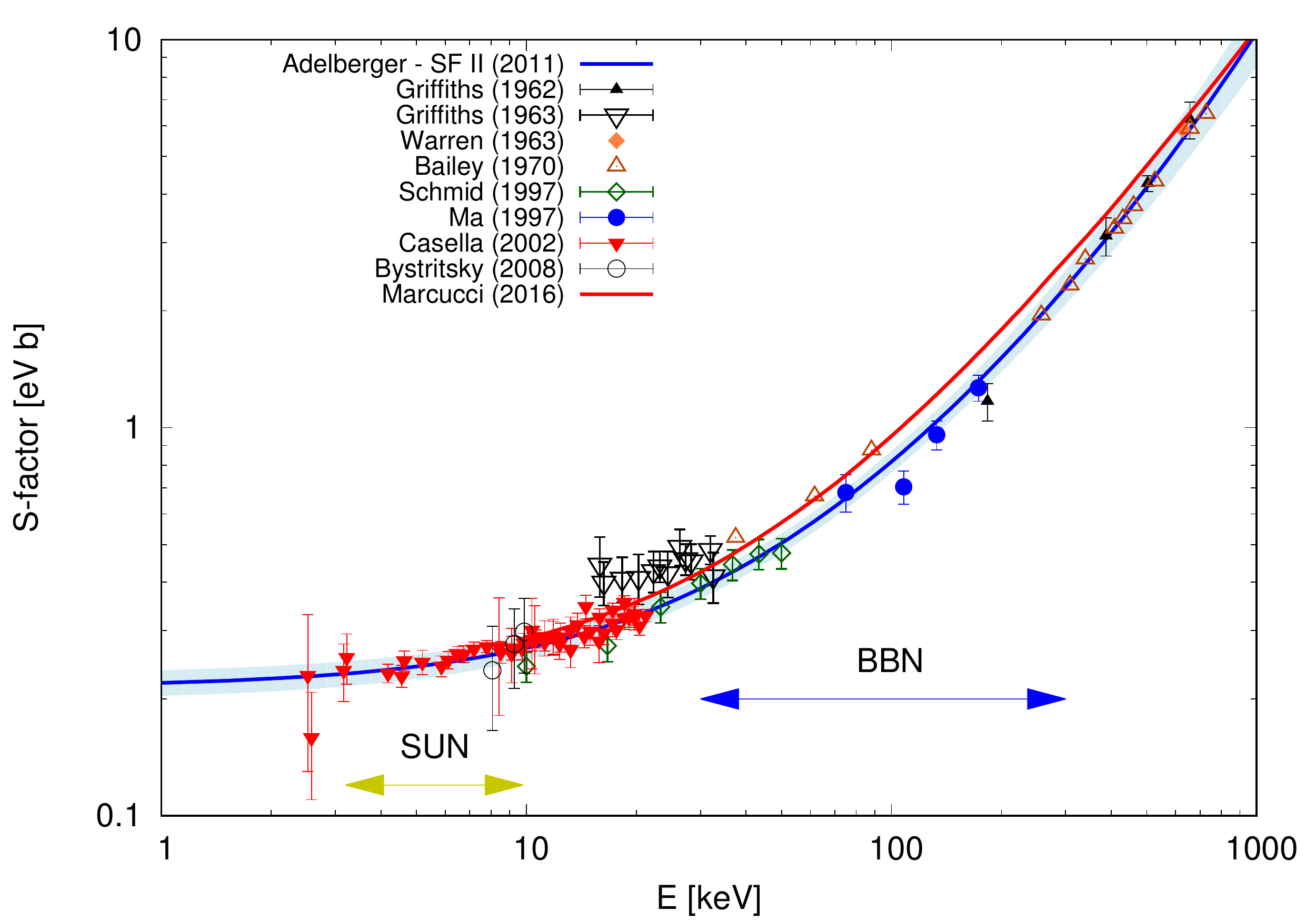}
  \caption{The $^2$H(p,$\gamma$)$^3$He astrophysical S-factor from different experiments \cite{Griffiths62-CJP,Griffiths63-CJP,Warren63-PR,Bailey70-CJP,Schmid97-PRC,Ma97-PRC,Bystritsky08-NIMA,Casella02-NPA}. In blue the Adelberger et al. fit
  (with the relative error band) \cite{Adelberger11-RMP}, in red the S-factor from ab-initio calculations \cite{Marcucci16-PRL}. The energy regions for the Gamow peak of the Sun and of BBN are
  marked.}
\label{pd}
    \end{center}
\end{figure}

Deuterium was the first isotope produced during BBN from the fusion of protons and neutrons. Its abundance is obtained from the spectroscopic measurement of metal-poor damped Lyman-alpha (LDA) systems. These systems are the oldest astrophysical environments where deuterium is detected. The value of [D/H] = (2.53 $\pm$ 0.04)$\times$10$^{-5}$ has been recently obtained \cite{Cooke14-ApJ} by averaging the best six astronomical measurements.
This value has to be compared with the one predicted by BBN \cite{DiValentino14-PRD}: [D/H] = (2.65 $\pm$ 0.07)$\times$10$^{-5}$.
As a consequence, deuterium is the only primordial isotope with a smaller uncertainty on the measurement than on the prediction.
This is due to the uncertainty on the cross section of $^2$H(p,$\gamma$)$^3$He. As a matter of fact, only a few experimental points exist in the energy region of interest with an overall systematic uncertainty at the 6-10\% level \cite{Adelberger11-RMP} (see Fig.\ref{pd}).

In particular, a higher value of the $^2$H(p,$\gamma$)$^3$He cross section in the 30-300 keV energy range would be required to obtain the observed deuterium abundance. Recent  ab initio calculations of the cross section have exactly provided such an enhancement \cite{Marcucci16-PRL}.

LUNA has already measured the cross section of $^2$H(p,$\gamma$)$^3$He (Q-value: 5.5 MeV) with the 50 kV accelerator \cite{Casella02-NPA} to explore the low energy region where the reaction takes place in pre-main sequence stars (burning the primordial deuterium at temperatures of about 1 MK) and in the Sun. A new experiment is now running at the 400 kV accelerator to cover the higher energy region of BBN.
Windowless gas targets filled with deuterium at 0.3 mbar constant pressure are used for the two phases of the experiment with different gamma ray detectors: high efficiency BGO and high resolution HPGe.
The high efficiency of the BGO ($\sim$70$\%$) reduces the dependence of the detector response on the angular distribution of the emitted $\gamma$ ray. On the other hand, during the second phase of the experiment the angular distribution will be inferred by exploiting the high energy resolution of the germanium detector and the Doppler effect influencing the energy of the produced $\gamma$ rays. With both detectors, thanks to the accelerator high current, a 1$\%$ statistical error is achievable, whereas
an overall systematic uncertainty of 3\% at most is aimed at.

With such an uncertainty on the cross section LUNA would be able either to confirm the agreement between astronomical observations and BBN calculations or to open a new scenario where non-standard cosmological models have to be considered \cite{DiValentino14-PRD}.
Moreover, a high precision measurement of the $^2$H(p,$\gamma$)$^3$He cross section at low energy will provide essential information for future ab-initio calculations.

\subsection{Lithium problems}\label{sec:lithium}

Lithium, with its two isotopes $^6$Li and $^7$Li, is the heaviest BBN element whose primordial abundance can be estimated with astronomical observations. Its abundance can be inferred from the absorption spectra of old stars. However, astrophysical processes like cosmic-ray interactions with the interstellar and intergalactic gas as well as neutrino reactions in supernovae and $^3$He burning in AGB stars can modify the abundance of lithium.
The only way to disentangle BBN nucleosynthesis from successive astrophysical processes is the measurement of lithium abundances in metal-poor halo stars in our Galaxy as a function of their metallicity \cite{Cyburt16-RMP}. The lower is the stellar metallicity the less is the dependence of lithium abundance from non-BBN nucleosynthesis processes. In particular, main sequence stars with temperature higher than about 6000 K show a constant lithium abundance \cite{Spite82-Nature}(the so called Spite plateau). Such an abundance, given by the strength of the $^7$Li absorption line at 670.7 nm, is taken as the primordial $^7$Li abundance. Its value is a factor 2-4 lower than the predicted one: such a discrepancy is known as the lithium problem (see table \ref{tab:abundances}).

\begin{figure}[h!]
    \begin{center}
  \includegraphics[scale=1.2]{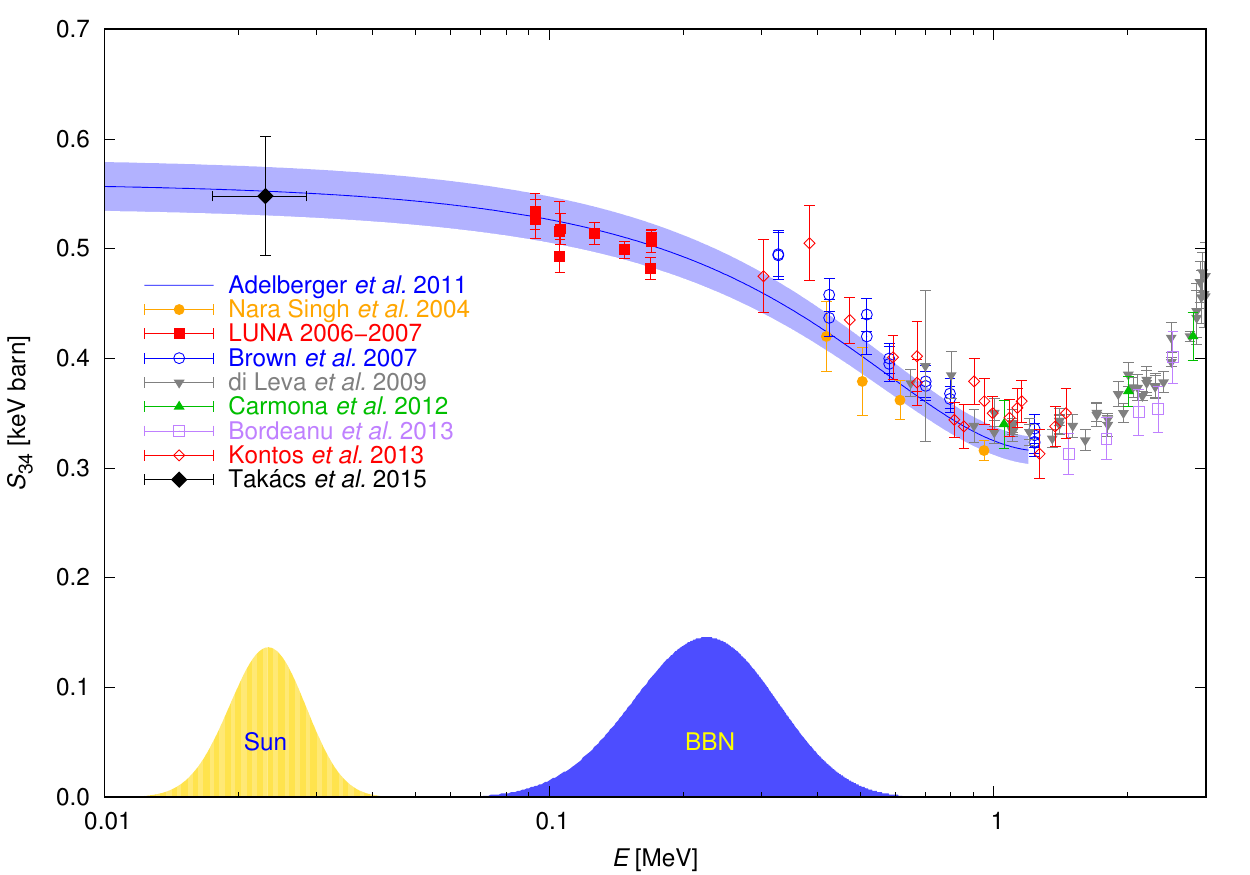}
  \caption{The $^3$He($\alpha$,$\gamma$)$^7$Be astrophysical S-factor from modern experiments \cite{NaraSingh04-PRL,Brown07-PRC,DiLeva09-PRL,Carmona12-PRC,Bordeanu13-NPA,Kontos13-PRC} and from the measurement of the solar neutrino flux \cite{Takacs15-PRD}. A theoretical curve rescaled to match the modern data \cite{Adelberger11-RMP} is shown. The Gamow peak for solar or BBN burning is depicted.}
\label{3He}
    \end{center}
\end{figure}

A nuclear physics solution to the $^7$Li problem is highly improbable. $^7$Li comes from the electron capture decay of $^7$Be, produced in the fusion reaction $^3$He($\alpha$,$\gamma$)$^7$Be.
The cross section of the reaction
can be determined either from the detection of
the prompt $\gamma$ rays
or from the counting of the decaying $^{7}$Be nuclei.
The latter requires the detection of the 478 keV $\gamma$ due to the excited $^{7}$Li populated in the decay of $^{7}$Be
(half-life: 53.22 days).
Both methods have been used in the past
but the S$_{3,4}$  extracted from the measurements
of the
induced $^{7}$Be activity was 13$\%$
higher than the one obtained
from the prompt $\gamma$-rays
\cite{Adelberger98-RMP}.

In the first phase of the experiment performed at LUNA, the $^{3}$He($\alpha$,$\gamma$)$^{7}$Be cross section has been obtained
from the activation data \cite{Bemmerer06-PRL,Gyurky07-PRC} alone.
In the second phase, a new high accuracy measurement using simultaneously
prompt and activation methods was performed down to the center of mass energy of 93 keV.
The prompt capture $\gamma$-ray was detected by a 135$\%$
germanium detector heavily shielded and placed in close geometry with the target.
The astrophysical factor obtained with the two methods \cite{Confortola07-PRC} has been found to be compatible with a total error on the measured cross section squeezed down to
4$\%$ in the center of mass energy region 93-170 keV. Even if the BBN region of interest is at higher energies, it is clear from Fig.\ref{3He} that a factor 2-4 reduction of the cross section inside the BBN window is ruled out.
As a consequence, the lithium problem remains unsolved \cite{Barbagallo16-PRL}.

The situation is less clear for what concerns the measurement of the $^6$Li abundance, where the existence of the Spite plateau has not been confirmed. As the absorption line of $^6$Li  is slightly shifted towards a higher wavelength compared to $^7$Li, the abundance of $^6$Li  has to be derived from the shape analysis of the $^7$Li absorption line.
Primordial $^6$Li is predicted to be about 5 orders of magnitude less
abundant than $^7$Li. However, in about a dozen cases of surveyed metal-poor stars a  $^6$Li abundance 3 orders of magnitude higher than the predicted one has been reported \cite{Asplund06-ApJ, Asplund08-AIPCP}, giving rise to the so called second lithium problem.
Recently, many of the claimed $^6$Li detections have been debated \cite{Lind13-AA}, however the situation is still unclear
\cite{Fields11-ARNPS,Steffen12-LiC}. In any case, before introducing new post-primordial processes or physics beyond the standard model
\cite{Fields11-ARNPS} it is mandatory to put the nuclear physics of $^6$Li production in standard BBN on solid experimental grounds. As a matter of fact, the reaction responsible for the
$^6$Li production, $^2$H($\alpha$,$\gamma$)$^6$Li,
has never been measured in the BBN energy window (30 - 300 keV). At these energies the reaction proceeds either via electric dipole (E1) or electric quadrupole (E2) direct
capture to the ground state of $^6$Li, in either case emitting a single $\gamma$ ray (Q-value=1.47 MeV). The E1 transition is strongly suppressed by selection rules
since E1 transitions between T=0 states are forbidden.

\begin{figure}[h!]
    \begin{center}
  \includegraphics[scale=1.2]{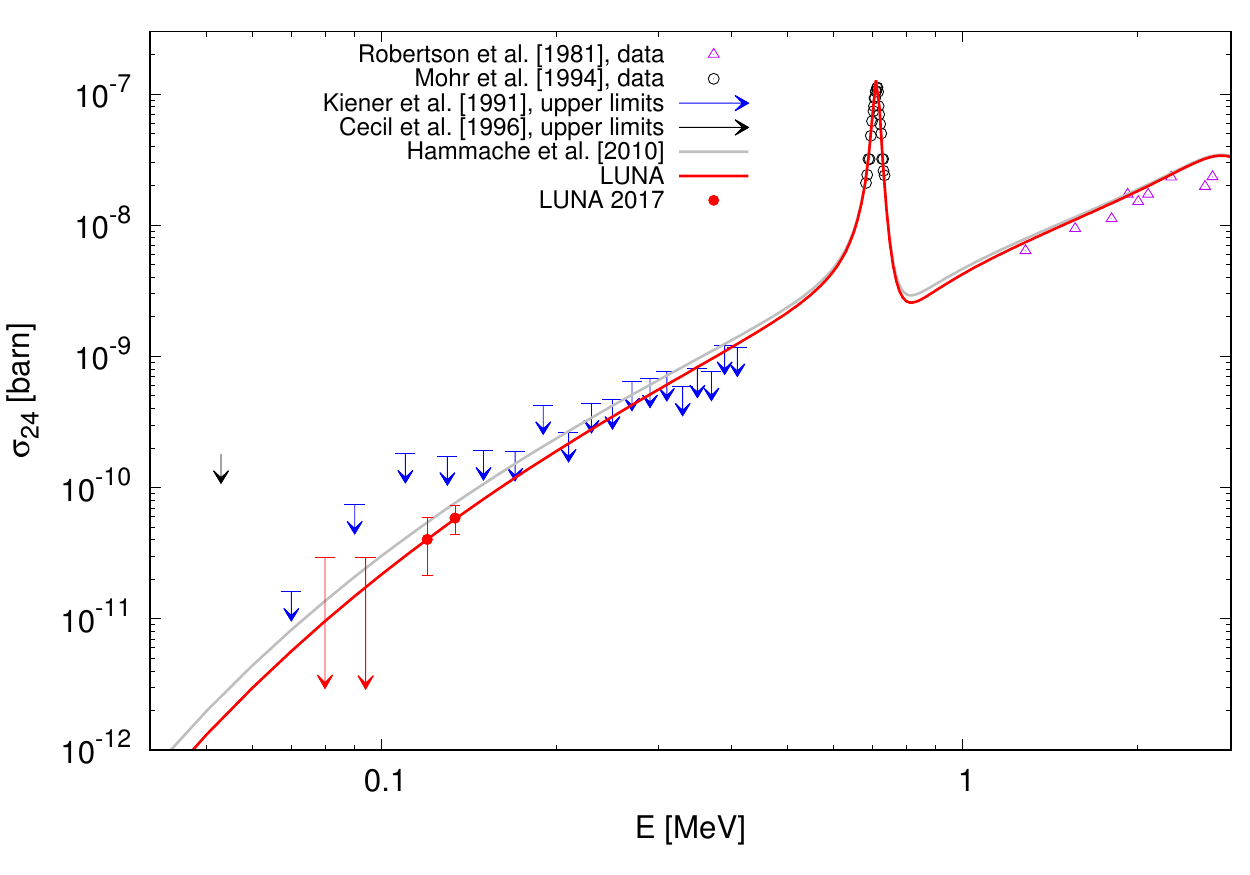}
  \caption{Cross section of the $^2$H($\alpha$,$\gamma$)$^6$Li reaction. The LUNA data \cite{Trezzi17-APP} are reported in red. Previous measurements and upper limits are
also reported: violet triangles \cite{Robertson81-PRL}, black circles \cite{Mohr94-PRC}, black arrows \cite{Kiener91-PRC} (upper limits), blue arrows \cite{Cecil96-PRC} (upper limits).  The LUNA recommended total cross section curve is given by the red full line \cite{Trezzi17-APP}. The Hammache et al. \cite{Hammache10-PRC} total cross section curve is also reported (grey full line).}
\label{dalpha}
    \end{center}
\end{figure}

LUNA has been the first experiment able to measure the $^2$H($\alpha$,$\gamma$)$^6$Li cross section inside the BBN energy range \cite{Anders14-PRL,Trezzi17-APP}.
The experimental setup consisted of a windowless gas target filled with 0.3 mbar of deuterium and a large HPGe detector placed at 90$^\circ$ angle with respect to the helium beam direction, in very close geometry (typical current 0.3 mA) \cite{Anders13-EPJA}.
The experiment had an important beam induced background due to energetic deuterons from elastic scattering of the $^4$He$^{+}$  beam on deuterium. These deuterons produce then
neutrons via the $^2$H(d,n)$^3$He reaction (Q-value=3.267 MeV). Subsequent inelastic neutron scattering in the germanium detector and in the shielding and structural materials
give rise to Compton background in the $^2$H($\alpha$,$\gamma$)$^6$Li region of interest. Despite this background, LUNA has measured the
$^2$H($\alpha$,$\gamma$)$^6$Li
cross section for the first time inside the BBN region of interest (Fig.\ref{dalpha}). Based on LUNA data an even lower $^6$Li/$^7 $Li ratio of (1.6$\pm$0.3)$\times$10$^{-5}$ is obtained \cite{Trezzi17-APP},
this way excluding a nuclear solution to the second lithium problem.

\section{Hydrogen burning and solar neutrinos}\label{sec:solar}

\begin{figure}[h]
    \begin{center}
  \includegraphics[scale=0.3]{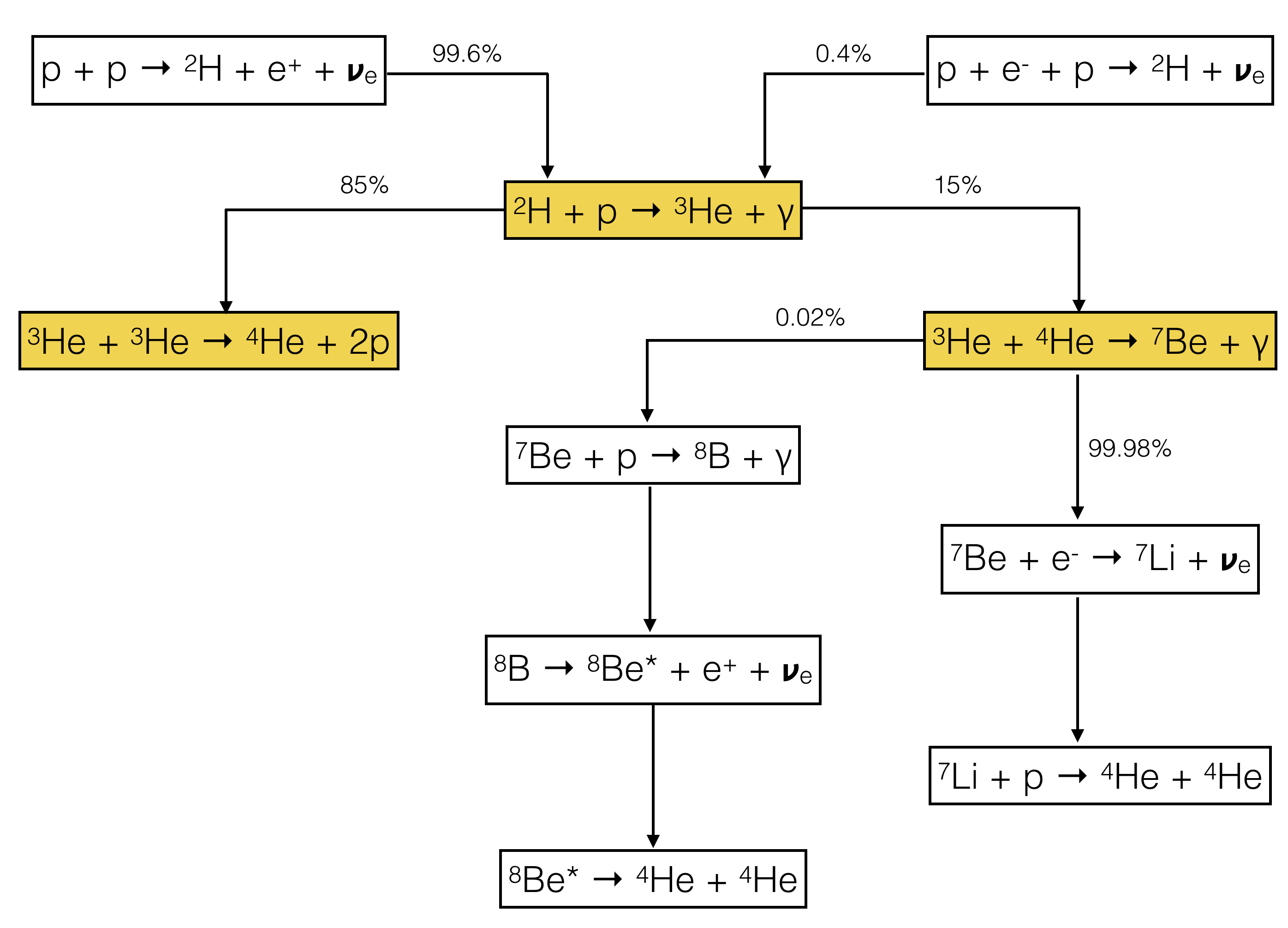}
  \caption{The proton-proton (pp) chain for hydrogen burning with the different branching ratios. The reactions
  studied by LUNA are highlighted.}
\label{ppchain}
    \end{center}
\end{figure}

The measurement of the $^{3}$He($^{3}$He,2p)$^{4}$He
cross section within the solar Gamow peak (16-27 keV) has been the reason why LUNA was born in 1991.
Such a reaction is a key one of the hydrogen burning proton-proton chain (Fig.\ref{ppchain}), which is responsible for more than 99$\%$ of the solar luminosity.
A resonance in its cross section at the thermal energy of the Sun was suggested at the beginning of the
seventies \cite{Fowler72-Nature,Fetisov72-PLB} to explain the low (as compared to
the Standard Solar Model predictions)
solar neutrino flux measured by the Homestake experiment \cite{homestake}. As a matter of fact, such a resonance would have decreased the
relative contribution of the alternative reaction
$^{3}$He($\alpha$,$\gamma$)$^{7}$Be, which generates the branch responsible for the production $^{7}$Be and $^{8}$B neutrinos interacting with
the  $^{37}$Cl targets of the Homestake experiment.

\begin{figure}[h]
    \begin{center}
  \includegraphics[scale=1.2]{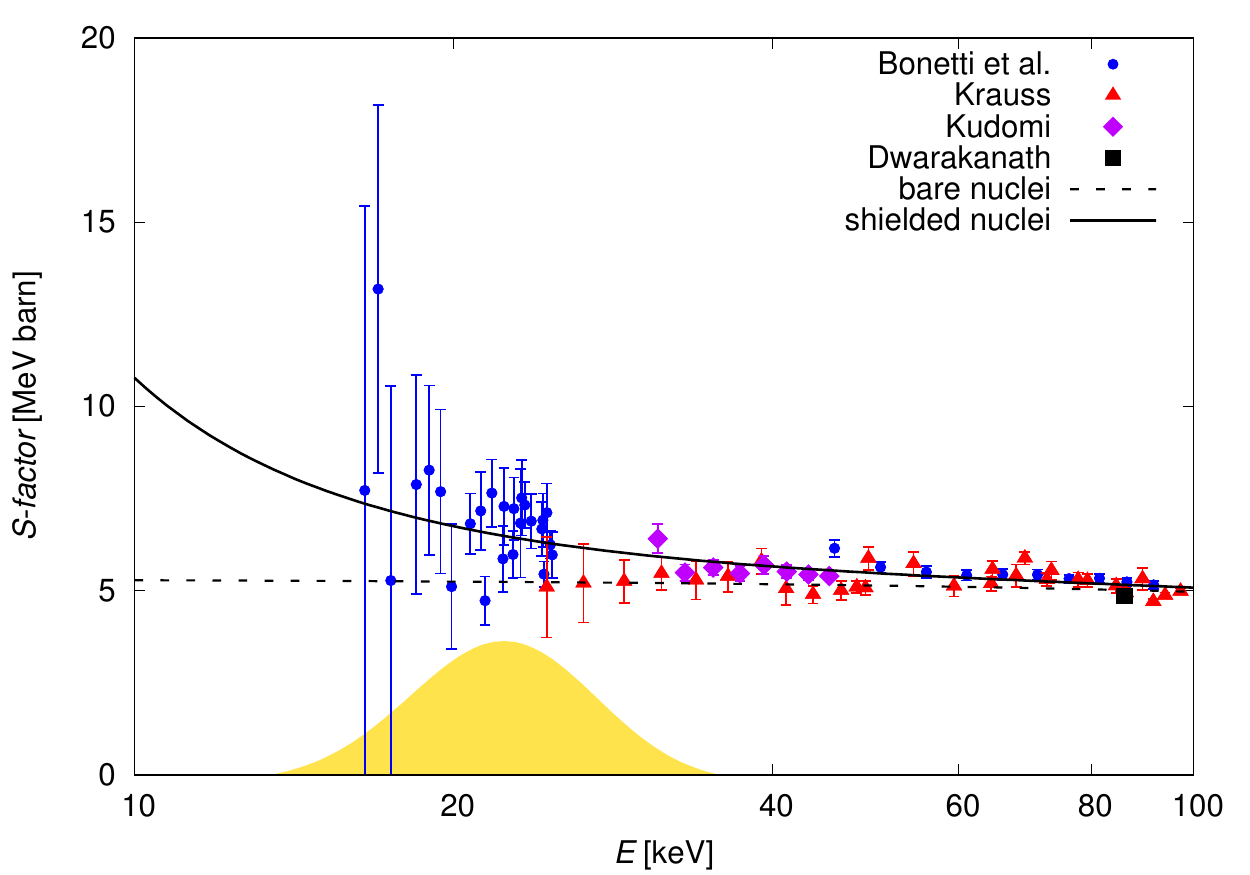}
  \caption{The astrophysical S-factor of $^{3}$He($^{3}$He,2p)$^{4}$He as function of energy. The lines are the fit to the astrophysical factors of bare and shielded nuclei.
  The solar Gamow peak is depicted.}
\label{3he+3he}
    \end{center}
\end{figure}

The final experimental set-up for the study of $^{3}$He($^{3}$He,2p)$^{4}$He was made of eight 1 mm thick silicon detectors of
5x5\,cm$^{2}$ area placed around the beam inside the windowless
target chamber filled with  $^{3}$He at the pressure of 0.5\,mbar.
The simultaneous detection of two protons has been the
signature which unambiguously identified the
$^{3}$He($^{3}$He,2p)$^{4}$He fusion reaction, thus completely suppressing the events due to
$^{3}$He(d,p)$^{4}$He which were the limiting background with the first detector set-up \cite{Junker98-PRC} (deuterium is contained in the $^{3}$He$^{+}$  beam as HD$^{+}$  molecule).
Fig.\ref{3he+3he}
shows the results from LUNA \cite{Bonetti99-PRL} together with
higher energy measurements \cite{Krauss87-NPA,Dwarakanath71-PRC,Kudomi04-PRC}. By fitting the observed energy dependence of the astrophysical factor we could also obtain an electron
screening potential $U_e$=294$\pm$47 eV, close to the one from the adiabatic limit (240 eV) \cite{Assenbaum87-ZPA}.

For the first time a
nuclear reaction has been measured in the laboratory at the
energy occurring in a star.
In particular, at the lowest energy of 16.5 keV the cross section is
0.02\,pb,
which corresponds to
a rate
of about 2 events/month, rather low even for the "silent"
experiments of underground physics.
No narrow resonance has been found and,
as a consequence, the astrophysical solution of the
$^{8}$B and $^{7}$Be solar neutrino problem based on
its existence was definitely ruled out.

$^{3}$He($\alpha$,$\gamma$)$^{7}$Be, the competing reaction for $^{3}$He burning, has also been measured by LUNA as described in the previous section. As a matter of fact, it is
the key reaction for the production of $^{7}$Be and $^{8}$B
neutrinos in the Sun.

\subsection{The metallicity of the Sun}

\begin{figure}[htb]
    \begin{center}
  \includegraphics[scale=0.5]{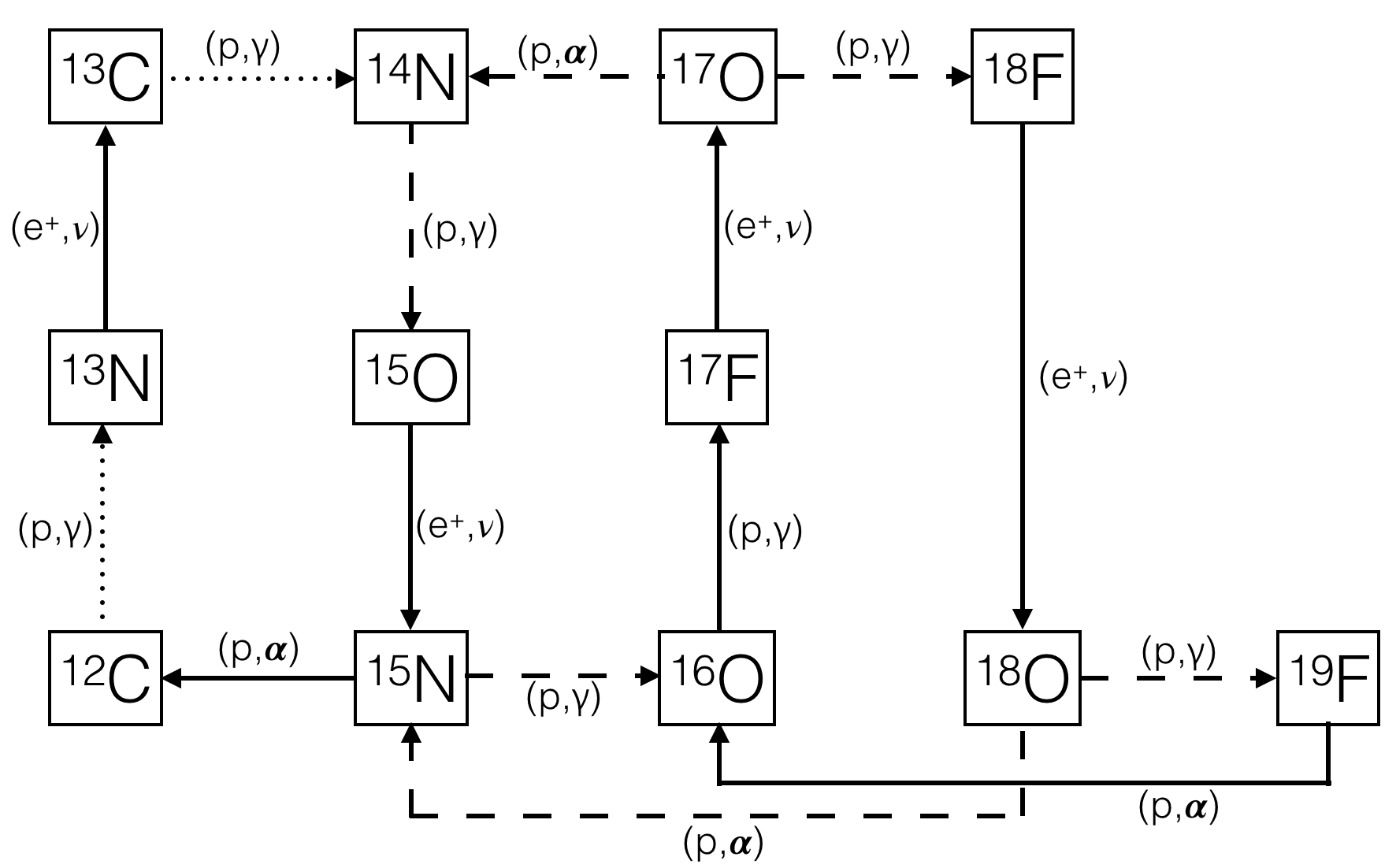}
  \caption{The CNO cycles. The reactions
  studied by LUNA up to now are highlighted by dashed arrows while possible future ones are represented by dotted lines.}
\label{cno}
    \end{center}
\end{figure}

$^{14}$N(p,$\gamma$)$^{15}$O
is  the slowest reaction
of the first CNO cycle (Fig.\ref{cno}) and it rules its energy production rate.
In particular, it is
the key reaction to predict
the $^{13}$N and $^{15}$O solar neutrino flux, which
depends almost linearly on its cross section.
In the first phase of the LUNA study, data have been obtained
down to 119 keV energy with
solid targets of TiN and a
germanium detector (the solar Gamow peak is between 20 and 35 keV).

\begin{figure}[htb]
    \begin{center}
  \includegraphics[scale=1.2]{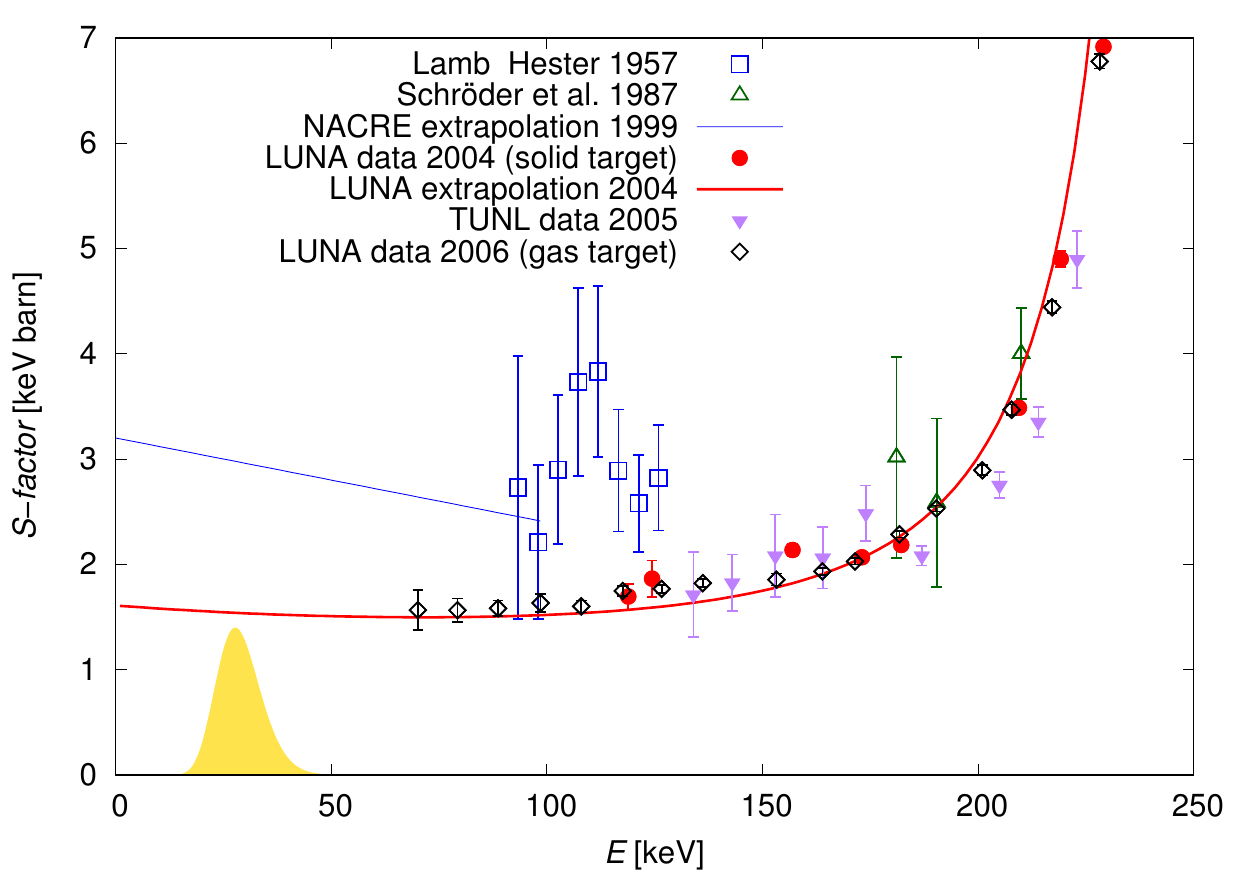}
  \caption{The astrophysical S-factor of $^{14}$N(p,$\gamma$)$^{15}$O as function of energy. The Gamow peak for solar burning is depicted.}
\label{p+14n}
    \end{center}
\end{figure}

This way, the five different radiative capture
transitions which contribute to the $^{14}$N(p,$\gamma$)$^{15}$O
cross section at low energy were measured.
The total cross section was then studied
down to very low energies in the second phase of the experiment
by using the 4$\pi$ BGO summing detector
placed around a
windowless gas target filled with nitrogen at 1 mbar pressure. At the lowest center of mass energy of 70 keV  a cross section of 0.24 pb was measured, with an event rate of
11 counts/day from the reaction.
The results (Fig.\ref{p+14n}) obtained first with  the germanium detector \cite{Formicola04-PLB,Imbriani05-EPJA}
and then with
the BGO set-up \cite{Lemut06-PLB} were about a factor of two lower than the existing extrapolation
\cite{Adelberger98-RMP,NACRE99-NPA}
from previous data \cite{Lamb57-PR,Schroeder87-NPA}
at very low energy. On the other hand, they were in good agreement with the reanalysis \cite{Angulo01-NPA} of \cite{Schroeder87-NPA} and with the results obtained with indirect methods
\cite{Mukhamedzhanov03-PRC}.
Because of this reduction the CNO neutrino yield
in the Sun is decreased by about a factor of two.

The lower cross section is affecting also stars which are more
evolved than our Sun. In particular, the lower limit on the age of the Universe inferred from the age of the oldest stellar populations,
the globular clusters, is increased by 0.7-1 billion years \cite{Imbriani04-AA}
up to 14 billion years, and thermal pulses
during the evolution of asymptotic giant branch (AGB) stars become more powerful, making
the dredge-up of
carbon to the surface more efficient \cite{Herwig04-ApJL}.
As a matter of fact, the luminosity of the turn-off point in the Hertzsprung-Russel diagram of a globular cluster, i.e. the point
where the main sequence turns toward cooler and brighter stars, is used to determine the age of the cluster and to derive a lower limit on the age of the Universe
\cite{Krauss03-Science}.
A star
at the turn-off point is burning hydrogen through the CNO
cycle, this is the reason why the $^{14}$N(p,$\gamma$)$^{15}$O cross section plays a key role
in the age determination.

In order to provide more precise data for the ground state capture, the most difficult one to be measured because of the summing problem,
we performed a third phase of
the  $^{14}$N(p,$\gamma$)$^{15}$O study with a composite germanium detector obtaining S$_{GS}$(0)=0.20$\pm$0.05 keVb \cite{Marta08-PRC}.
This way the total error on the extrapolated S-factor has been
reduced to 8$\%$:
S$_{1,14}$(0)=1.57$\pm$0.13 keVb. The LUNA result is in agreement with the recommended one given in the 2011 compilation of solar fusion
cross sections \cite{Adelberger11-RMP}: S$_{1,14}$(0)=1.66$\pm$0.12 keVb.
It is important to have such a relatively small error (we started with a 50$\%$ one \cite{Adelberger98-RMP}) since,
finally solved the solar neutrino problem, we are now facing the solar composition problem:
the conflict between the helioseismology results and the predictions of the Standard Solar Model once the new value of the metal abundance (i.e. the amount of elements different from hydrogen and helium) that emerged from improved modeling
of the solar photosphere are used \cite{Serenelli16-EPJA}.

In the near future we will have the first measurement of the CNO neutrinos from the Sun \cite{Ianni17-PPNP}. At that moment the carbon and nitrogen content of the Sun core
will be obtained
from the comparison between the measured CNO neutrino flux and the predicted one \cite{Serenelli16-EPJA}.
As a matter of fact, the CNO neutrino flux depends linearly both on the $^{14}$N(p,$\gamma$)$^{15}$O cross section and on
the carbon and nitrogen abundance and it decreases by about 30$\%$ in going from
the high to the low metallicity scenario. Finally, the value of the core metallicity will allow to test
whether the early Sun was chemically homogeneous, a
key assumption of the Standard Solar Model \cite{Haxton08-ApJ}.

\subsubsection{A new study of $^{14}$N(p,$\gamma$)$^{15}$O}

A new result \cite{Li16-PRC} has been published last year on the measurement of $^{14}$N(p,$\gamma$)$^{15}$O
performed by using a 1 MV and a 4 MV accelerators at the University of Notre Dame. The cross
of $^{14}$N(p,$\gamma$)$^{15}$O
was measured over the proton energy range from 0.7 to 3.6 MeV for both the ground state and the 6.79 MeV transition.
Since the cross section
cannot be measured at the solar Gamow peak, where the rate would be too low, the unavoidable extrapolation with R-matrix fitting is
sensitive also to the cross section value in the MeV region. In particular, some inconsistencies have been found in \cite{Li16-PRC} for the ground state transition and,
as a consequence, a large systematic uncertainty is recommended: S$_{GS}$(0)=0.42$\pm$0.04(stat)$^{+0.09}_{-0.19}$(syst) keV b. It is now mandatory to check
this result by performing the cross section measurement with a single accelerator over a wide energy region (this way removing the problem of normalization
among different experiments). LUNA MV is perfectly suited for such a measurement, with the additional benefit of the Gran Sasso suppressed background.
Using several HPGe detectors positioned at different angles in order to study the angular dependence of
the cross section, it may be possible to connect to the existing low-energy LUNA 400 kV data and also gain additional information hitherto unavailable.

\section{AGB and Classical Novae}\label{sec:AGB}

A rich program of nuclear astrophysics
mainly devoted to CNO, Ne-Na and Mg-Al cycles started at the end of the solar phase of LUNA. Of particular interest are those bridge reactions which are connecting one cycle to the next, or which are key
ingredients of gamma astronomy.
Due to the higher Coulomb barrier of the reactions involved, the cycles become important at temperatures higher than the one of our Sun, i.e. during  hydrogen burning in the shell of AGB stars and during the thermonuclear runaway of classical
novae (about 30-100, and 100-400 million degrees, respectively).
Relatively unimportant for energy generation, these cycles are essential for the cooking of the light nuclei up to $^{27}$Al.

\subsection{The CNO cycles}

The first CNO cycle is linked to the second one by the  $^{15}$N(p,$\gamma$)$^{16}$O reaction, this way affecting the synthesis of $^{16}$O \cite{Iliadis02-ApJSS} in classical novae. Classical novae are binary systems made of a white dwarf (WD) and a low-mass companion star still on the main sequence. When the H-rich gas from the companion finally accumulates on the WD surface, it produces a semi degenerate envelope. The thermonuclear runaway, i.e. the explosive burning, is triggered inside the envelope by $^{12}$C(p,$\gamma$)$^{13}$N, followed by series of nuclear fusions.
During this process, the ratio between the $^{15}$N(p,$\gamma$)$^{16}$O rate and the much larger rate of $^{15}$N(p,$\alpha$)$^{12}$C determines
the leak frequency from the first to the second CNO cycle, which leads to the production of the oxygen isotopes \cite{Jose98-ApJ}.
While the (p,$\alpha$) channel was very well known, only two measurements of $^{15}$N(p,$\gamma$)$^{16}$O existed until 2009 \cite{Rolfs74-NPA,Hebbard60-NP}.
They were not in agreement between each other and only one of them
\cite{Rolfs74-NPA} was considered in the NACRE database to give an S-factor $S(0) = 64\pm6$~keV \cite{NACRE99-NPA}.
A different value, almost a factor of 2 lower than the previously adopted one, was coming from an indirect measurement based on the study of $^{15}$N($^3$He,d)$^{16}$O \cite{Mukh08-PRC}.

In LUNA it has been possible to measure the $^{15}$N(p,$\gamma$)$^{16}$O reaction in the energy region of classical novae.
The first results came from the analysis of data previously acquired for the study of the $^{14}$N(p,$\gamma$)$^{15}$O reaction with the high efficiency BGO setup \cite{Bemmerer06-NPA,Lemut06-PLB} and a windowless gas target filled with
natural nitrogen (0.4\% $^{15}$N).
A severe beam induced background  was due to the $\gamma$ rays from $^{11}$B(p,$\gamma$)$^{12}$C \cite{Cecil92-NPA,Bemmerer09-JPG}.
The second phase of the experiment was performed with solid TiN targets 96\% enriched  in $^{15}$N, this way enhancing the
signal to noise ratio.
With these targets it has been possible to measure down to 130 keV using a high purity germanium detector \cite{LeBlanc10-PRC} and then down to 70 keV with the  4$\pi$-BGO detector. This way, LUNA covered for the first time the Gamow window of nova explosion measuring a cross section which is  a factor of 2 lower than the NACRE extrapolation \cite{NACRE99-NPA}. As a consequence, the amount of expected $^{16}$O synthesized in nova explosion has been reduced by 30\% \cite{Caciolli11-AA}.

The two other key reactions involved in the production of oxygen isotopes in nova and AGB stars are $^{17}$O(p,$\gamma$)$^{18}$F and $^{17}$O(p,$\alpha$)$^{14}$N.
Their ratio determines the leak frequency from the second to the third CNO cycle and it affects not only the synthesis of $^{17}$O and  $^{18}$O, but also the ones of $^{18}$F and $^{19}$F.
In particular, $^{17}$O(p,$\gamma$)$^{18}$F produces $^{18}$F which decays with $t_{1/2}$ = 110 min to $^{18}$O and therefore contributes to the production of $^{19}$F and $^{15}$N due to the hydrogen fusion with $^{18}$O \cite{Jose98-ApJ,Jose07-ApJL}. Finally, we observe that the intensity of the expected gamma ray flux  at 511 keV from classical novae, due to $\beta^+$ annihilation from $^{18}$F, strongly depends on the rate of  $^{17}$O(p,$\gamma$)$^{18}$F and $^{17}$O(p,$\alpha$)$^{14}$N.

At low energies the $^{17}$O(p,$\gamma$)$^{18}$F cross section depends on the
narrow resonances at 64.5~keV and 183~keV, on the tail of two broad resonances at 557~keV and 677~keV and on the direct capture component (the strength of the 64.5 keV is too low to be measured and its value only comes from an indirect evaluation \cite{NACRE99-NPA}). The 183~keV resonance has been studied before LUNA by
either detecting the prompt gamma rays \cite{Fox04-PRL}  or by  detecting the 511~keV $\gamma$ due to the $\beta^+$ from the decay of $^{18}$F \cite{Chafa07-PRC} (activation method). The results from the two
experiments differ by more than 2 standard deviations.
The non resonant component of the cross section was obtained by several groups both in direct \cite{Rolfs73-NPA_SpecFac,Newton10-PRC,Kontos12-PRC,Buckner15-PRC} and inverse \cite{Hager12-PRC} kinematics but with quite large uncertainties.
LUNA measured the cross section over a wide energy range (167-370 keV) appropriate to hydrogen burning in classical novae both by detecting the prompt gamma rays with a germanium detector and by activation, using  Ta$_2$O$_5$ targets enriched in $^{17}$O (65\%) \cite{Caciolli12-EPJA}. The two methods gave results in quite good agreement between each other. The strength
$\omega \gamma = 1.67 \pm 0.12 \mu$eV has been obtained for the  $E$ = 183 keV resonance and the uncertainty on the cross section has been reduced by a factor of 4 with respect to previous experiments \cite{Scott12-PRL,DiLeva14-PRC}. As a consequence, the uncertainties on the predicted yields of oxygen and fluorine isotopes in classical novae have been reduced from 40-50\% down to
less than 10\% \cite{DiLeva14-PRC}.
\begin{figure}[ht]
    \begin{center}
  \includegraphics[angle = 0, scale=.8]{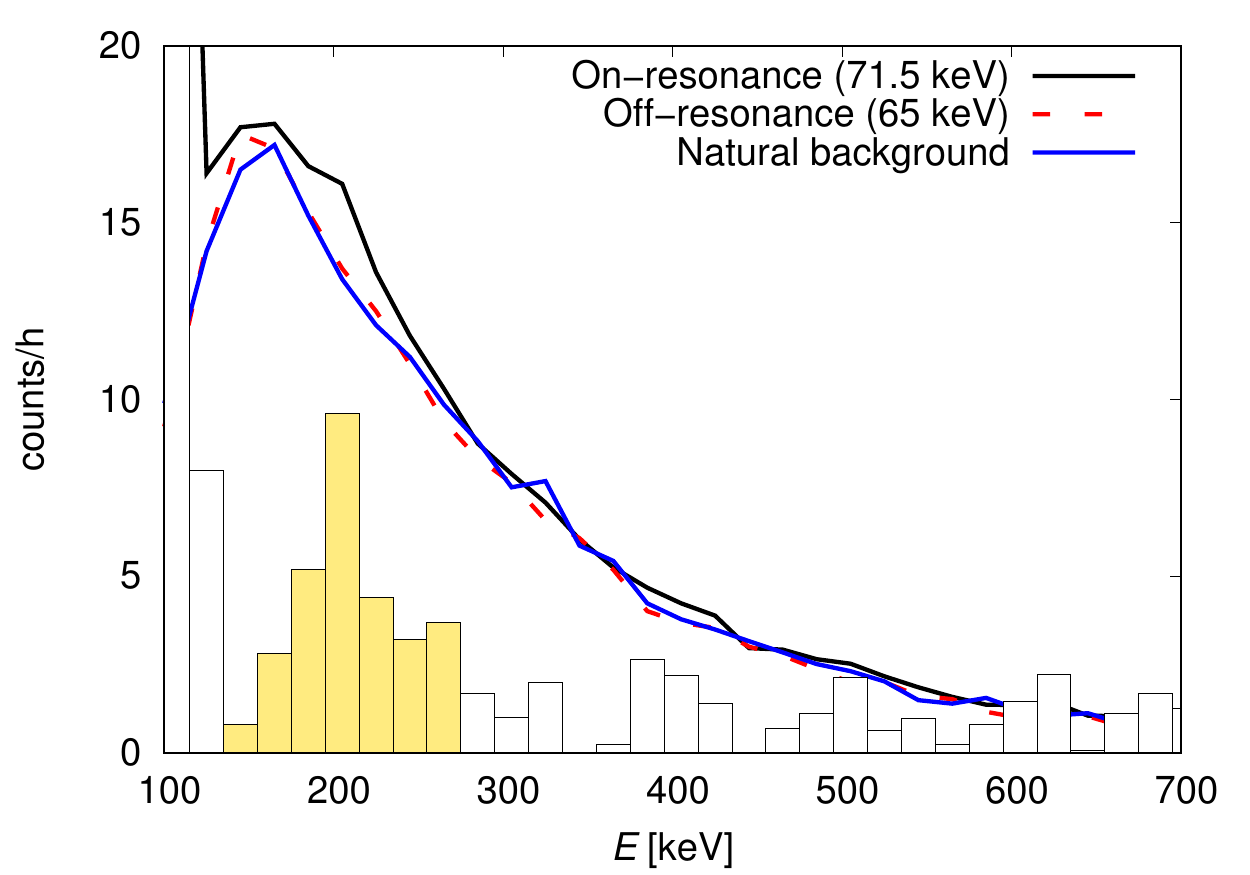}
  \caption{On-resonance (proton energy of 71.5 keV), off-resonance and background spectra time normalized for the study of the 64.5 keV resonance of
  $^{17}$O(p,$\alpha$)$^{14}$N \cite{Bruno16-PRL}. The histogram, here in arbitrary units, results from the difference between the on-resonance spectrum and the background one.}
\label{fig:bruno}
    \end{center}
\end{figure}

Alpha particle detectors running underground exhibit a small, about a factor of 15, but still crucial background suppression as compared to overground \cite{Bruno15-EPJA}. Thanks to this, not only the 183 keV resonance \cite{Bruno15-EPJA} but also the 64.5~keV one could be properly measured by LUNA in $^{17}$O(p,$\alpha$)$^{14}$N \cite{Bruno16-PRL}.
From the existing measurements \cite{Berheide92-ZPA,Niemeyer96-PhD,Blackmon95-PRL}, NACRE gave a strength of $\omega \gamma = 5.5^{+1.8}_{-1.5}$ neV \cite{NACRE99-NPA}.
Thanks to the background suppression achieved underground \cite{Bruno15-EPJA}, a clear signal from the resonance could be detected by an array of eight silicon detectors
(Fig.\ref{fig:bruno}). A strength $\omega \gamma = 10.0 \pm 1.4_{stat} \pm 0.7_{syst}$ neV was measured \cite{Bruno16-PRL}. Such a value is about a factor of 2 higher than the previously estimated, thus leading to a factor of 2 increase
in the reaction rate of shell hydrogen burning in red giant and asymptotic giant branch (AGB) stars.

In particular, the new resonance strength has increased by 20\% the predicted $^{16}$O/$^{17}$O ratio \cite{Straniero17-AA}
after the first dredge up in red giant stars and it has allowed for the firm identification of the production site of a star grain population \cite{Lugaro17-NatureAstronomy}.
Stardust grains recovered from meteorites are the survivors of the pre-solar dust and their isotopic abundances identify the stellar environment where they
are coming from. There has been a long-standing puzzle in stardust composition: intermediate mass stars of 4-8 solar masses are expected to produce stardust but with an isotopic
signature which has never been measured. The LUNA result now leads to
$^{17}$O$/^{16}$O ratios for shell hydrogen burning in AGB stars of 4-8 solar masses (60-80 MK) which very well agree with the one measured in 10-25\% of the oxygen rich grains.

\subsection{The NeNa cycle}

\begin{figure}[h!]
    \begin{center}
  \includegraphics[scale=0.48]{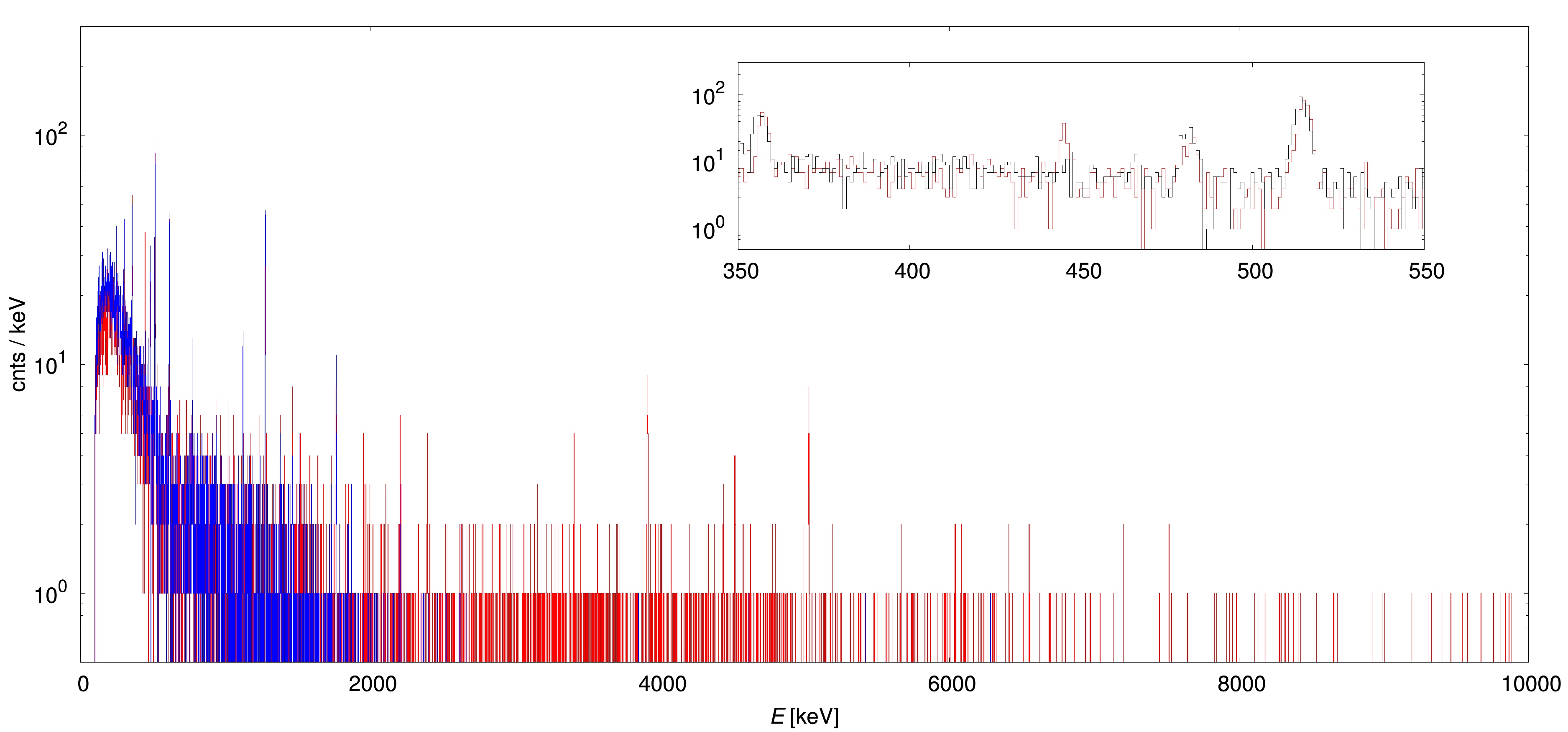}
  \caption{A spectrum acquired at the energy of the 156 keV resonance of the $^{22}$Ne(p,$\gamma$)$^{23}$Na reaction in red. In blue the spectrum acquired below the resonance is also shown normalised to the same integrated charge. In the inset, the 440 keV gamma line is clearly visible above the background.}
\label{fig:HPGE_spectrum_Ne}
    \end{center}
\end{figure}

$^{22}$Ne(p,$\gamma$)$^{23}$Na is the reaction of the Ne-Na cycle which had the highest uncertainty, up to a factor of 1000 at temperatures relevant for nuclesoynthesis
in AGB stars and classical novae. Its uncertainty is due to a large number of predicted but not yet observed resonances at low energy. A direct study of low energy
resonances has been performed at LUNA by delivering a high proton beam to a $^{22}$Ne gas target. Two high purity germanium detectors placed at 55$^{\circ}$ and 90$^{\circ}$
with respect to the beam direction revealed the prompt gamma rays. The detectors were enclosed in a copper and lead shield to suppress the laboratory background
\cite{Cavanna14-EPJA}. Three resonances at 156.2, 189.5 and 259.7 keV proton beam energy have been observed for the first time \cite{Cavanna15-PRL} (Fig.\ref{fig:HPGE_spectrum_Ne}). The $\gamma$-decay branching ratios of the corresponding energy levels in $^{23}$Na
have also been precisely measured \cite{Depalo16-PRC}. For the strength of three additional tentative resonances at 71, 105 and 215 keV only new and more stringent upper limits could be derived.

\begin{figure}[h!]
    \begin{center}
  \includegraphics[scale=1.0]{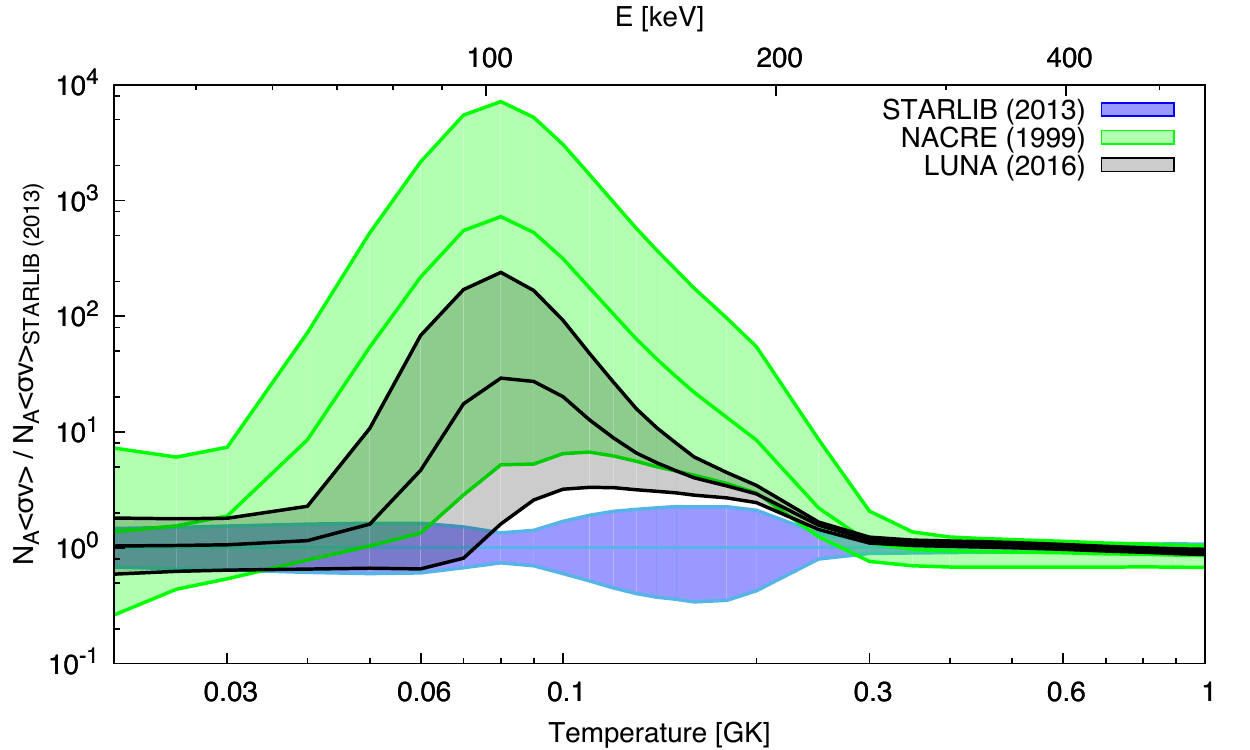}
  \caption{The reaction rate of LUNA for $^{22}$Ne(p,$\gamma$)$^{23}$Na plotted together with the STARLIB and NACRE ones and normalized to the STARLIB rate.}
\label{fig:22Ne_RelativeRate}
    \end{center}
\end{figure}

Finally, an updated reaction rate has been obtained, which is significantly higher than the one given in the most recent compilation of reaction rates \cite{starlib}
(Fig.\ref{fig:22Ne_RelativeRate}).
As a consequence, new values for the ejected mass of $^{22}$Ne and $^{23}$Na in thermally pulsing AGB stars have been obtained with much reduced uncertainties \cite{Slemer17-MNRAS}.

The remaining uncertainty on the cross section at low energy is due to the possible resonances at 71 and 105 keV. In order to reduce this uncertainty, a new measurement has been performed at LUNA by using the high efficiency 4$\pi$-BGO detector. The data analysis is still ongoing. With this data it will be possible to measure the direct capture contribution and either to measure the strength of the two resonances or to give upper limits which make negligible their contribution to hydrogen burning in AGB stars.

\subsection{The MgAl cycle}

The MgAl cycle plays a relevant role in the synthesis of Mg and Al isotopes and it is activated in the H-burning regions of stars at temperatures higher than 30-40 MK.
$^{25}$Mg(p,$\gamma$)$^{26}$Al is a key reaction of the cycle: it proceeds either to the ground state of $^{26}$Al (60-80\% probability, t$_{1/2} \sim 7 \cdot 10^5$ yr) or to the 228 keV isomeric state (t$_{1/2}$=6 s). The $^{26}$Al ground state decays  into the first excited state of $^{26}$Mg giving rise to the 1.809 MeV
$\gamma$ line, one of the few seen in the $\gamma$ sky of the Milky Way \cite{COMPTEL,INTEGRAL}(its detection has been a prove that nucleosynthesis is a continuing process). On the contrary, the $^{26}$Al isomeric state
decays to the ground state of $^{26}$Mg and does not produce any $\gamma$ ray.

LUNA at first measured the 304 keV resonance \cite{Limata10-PRC} and then studied the strengths of the 92 and 189 keV resonances with the 4$\pi$-BGO detector \cite{Strieder12-PLB}. In addition, the 189 keV resonance was also studied with a germanium detector
in order to precisely determine the branching ratios. In particular, the strength  $\omega \gamma$ = (2.9$\pm$0.6)$\cdot$10$^{-10}$ eV has been measured for the 92 keV resonance, probably
the lowest ever measured resonance strength \cite{Strieder12-PLB}. The reaction rate for temperatures between 50 and 150 MK obtained by LUNA is a factor of 2 higher than the one previously adopted \cite{NACRE99-NPA,Iliadis10-NPA841_31},  while the production rate of the isomeric state is up to a factor of 5 larger \cite{Straniero13-ApJ}.
As a consequence, the expected production of $^{26}$Al ground state in stellar H-burning zones is lower than previously estimated, confirming core collapse supernovae as the main source of $^{26}$Al in the Milky Way \cite{Iliadis11-ApJSS}.

\subsection{Future  potential at LUNA 400}

The reactions till now studied underground by LUNA are summarized in Table \ref{tab:past}.
Data have already been taken and the analysis is going on for $^{23}$Na(p,$\gamma$)$^{24}$Mg, $^{18}$O(p,$\gamma$)$^{19}$F and $^{18}$O(p,$\alpha$)$^{15}$N, with the first connecting the NeNa cycle to the MgAl one
and the second, followed by  $^{19}$F(p,$\gamma$)$^{20}$Ne, connecting the CNO cycles to the NeNa one.

In addition to $^2$H(p,$\gamma$)$^3$He, previously discussed in connection with BBN, LUNA is now measuring $^{6}$Li(p,$\gamma$)$^{7}$Be, to verify the possible existence of a resonance at 195 keV \cite{He13-PLB}.
Other reactions offering great potential for study in the next years, in addition to $^{13}$C($\alpha$,n)$^{16}$O  which is described in the next chapter, are
$^{22}$Ne($\alpha$,$\gamma$)$^{26}$Mg, the reaction competing with $^{22}$Ne($\alpha$,n)$^{25}$Mg, and the two reactions which determine the relative abundance of $^{12}$C and $^{13}$C in the deepest layers of the H-rich envelope of any star:
$^{12}$C(p,$\gamma$)$^{13}$N and $^{13}$C(p,$\gamma$)$^{14}$N.

\begin{table}[h]\footnotesize
\begin{center}

    \begin{tabular}{| c | c | c | c | c | c |}
    \hline
     \textbf{Reaction} & \textbf{Burning Stage} & {\bf Range $E$ [keV]} & \textbf{Target} & {\bf Detector} & \textbf{Reference} \\
    \hline
    \hline
    $^3$He($^3$He,2p)$^4$He & pp-chain & 16.5 - 24.4 & gas & Si & \cite{Arpesella96-PLB,Junker98-PRC,Bonetti99-PRL} \\
    $^2$H($^3$He,p)$^4$He & pp / $e^-$ screening & 5.4 - 31.3 & gas & Si & \cite{Prati94-ZPA,Costantini00-PLB,Zavatarelli01-NPA} \\
    $^2$H(p,$\gamma$)$^3$He & pp-chain/BBN & 2.5 - 22 & gas &  4$\pi$-BGO & \cite{Casella02-NPA} \\
    $^3$He($\alpha$,$\gamma$)$^7$Be & pp-chain/BBN & 93 - 170 & gas &  HPGe & \cite{Bemmerer06-PRL,Gyurky07-PRC,Confortola07-PRC,Costantini08-NPA} \\
    $^2$H($\alpha$,$\gamma$)$^6$Li & BBN & 80 - 133 & gas &  HPGe & \cite{Anders13-EPJA,Anders14-PRL,Trezzi17-APP} \\
    $^{14}$N(p,$\gamma$)$^{15}$O & CNO & 70 - 228 & gas &  4$\pi$-BGO & \cite{Bemmerer06-NPA,Lemut06-PLB} \\
     &  & 119 - 370 & solid &  HPGe & \cite{Formicola04-PLB,Imbriani04-AA,Imbriani05-EPJA,Marta08-PRC,Marta11-PRC} \\
    $^{15}$N(p,$\gamma$)$^{16}$O & CNO & 90 - 230 & gas &  4$\pi$-BGO & \cite{Bemmerer09-JPG} \\
     &  & 70 - 375 & solid &  HPGe/4$\pi$-BGO & \cite{LeBlanc10-PRC,Caciolli11-AA} \\
    $^{17}$O(p,$\gamma$)$^{18}$F & CNO & 167 - 370 & solid &  HPGe  & \cite{Scott12-PRL,DiLeva14-PRC} \\
    $^{17}$O(p,$\alpha$)$^{14}$N & CNO & 64.5, 183 (R) & solid &  Si & \cite{Bruno15-EPJA,Bruno16-PRL,Lugaro17-NatureAstronomy,Straniero17-AA} \\
    $^{18}$O(p,$\alpha$)$^{15}$N & CNO & 143 (R) & solid &  Si & \cite{Bruno15-EPJA} \\
    $^{22}$Ne(p,$\gamma$)$^{23}$Na & Ne-Na & 156.2,189.5,259.7 (R) & gas &  HPGe & \cite{Cavanna14-EPJA,Cavanna15-PRL,Depalo16-PRC,Slemer17-MNRAS} \\
    &  & 70,105,215 (R) & gas &  HPGe & \cite{Cavanna15-PRL,Depalo16-PRC,Slemer17-MNRAS} \\
    $^{24}$Mg(p,$\gamma$)$^{25}$Al & Mg-Al & 214(R) & solid & HPGe/4$\pi$-BGO & \cite{Limata10-PRC} \\
    $^{25}$Mg(p,$\gamma$)$^{26}$Al & Mg-Al & 92,130,189.5,304 (R) & solid & HPGe/4$\pi$-BGO & \cite{Limata10-PRC,Strieder12-PLB,Straniero13-ApJ} \\
    $^{26}$Mg(p,$\gamma$)$^{27}$Al & Mg-Al & 326 (R) & solid & HPGe/4$\pi$-BGO & \cite{Limata10-PRC} \\

       \hline
    \end{tabular}
\caption{The reactions studied underground by LUNA (May 2017) and the relative references. The resonance energy (R) is
given in the center of mass system.}
\label{tab:past}
\end{center}
\end{table}
\newcommand{\can}{$^{13}$C($\alpha$,n)$^{16}$O}

\section{The neutrons for the s-process}\label{sec:LUNAMV}

The neutrons for the astrophysical s-process are supplied by two ($\alpha$,n) reactions: At low temperatures, $\sim$ 90 MK, $^{13}$C($\alpha$,n)$^{16}$O plays the major role
\cite{Gallino98-ApJ,Lugaro03-ApJ}. It operates in the He-rich shell of low-mass (less than 4 solar masses) AGB stars, the so called intershell, between the helium and hydrogen burning shells. It is the neutron source reaction that allows the creation of the majority of the s-process elements.
At higher temperatures, $\sim$ 300 MK, $^{22}$Ne from one or two helium capture processes on $^{18}$O or $^{14}$N is the seed for the $^{22}$Ne($\alpha$,n)$^{25}$Mg reaction.
$^{22}$Ne($\alpha$,n)$^{25}$Mg operates in the He-burning shell of high-mass (more than 4 solar masses) AGB stars and during the core-He burning and the shell-C burning of massive stars (more than 10 solar masses). The latter is responsible for the synthesis of
the s-process elements with mass number A smaller than 90.

\subsection{$^{13}$C($\alpha$,n)$^{16}$O}\label{subsec:c13an}

The cosmic creation of roughly half of all elements heavier than iron, including
metals, such as W and Pd, as well as rare earth, such
as La and Nd, occurs in AGB stars thanks to the neutrons from  $^{13}$C($\alpha$,n)$^{16}$O. The rate of this reaction determines whether the $^{13}$C nuclei burn in radiative conditions between two convective pulses or, instead, they are ingested in the convective thermal pulses driven by He burning.
If $^{13}$C is ingested in the convective region, the \can\ reaction becomes also an energy source, which affects the development of the thermal pulse itself \cite{Bazan93-ApJ,Cristallo09-PASA}.

\begin{figure}[t!]
\begin{center}
\includegraphics[width=18 cm]{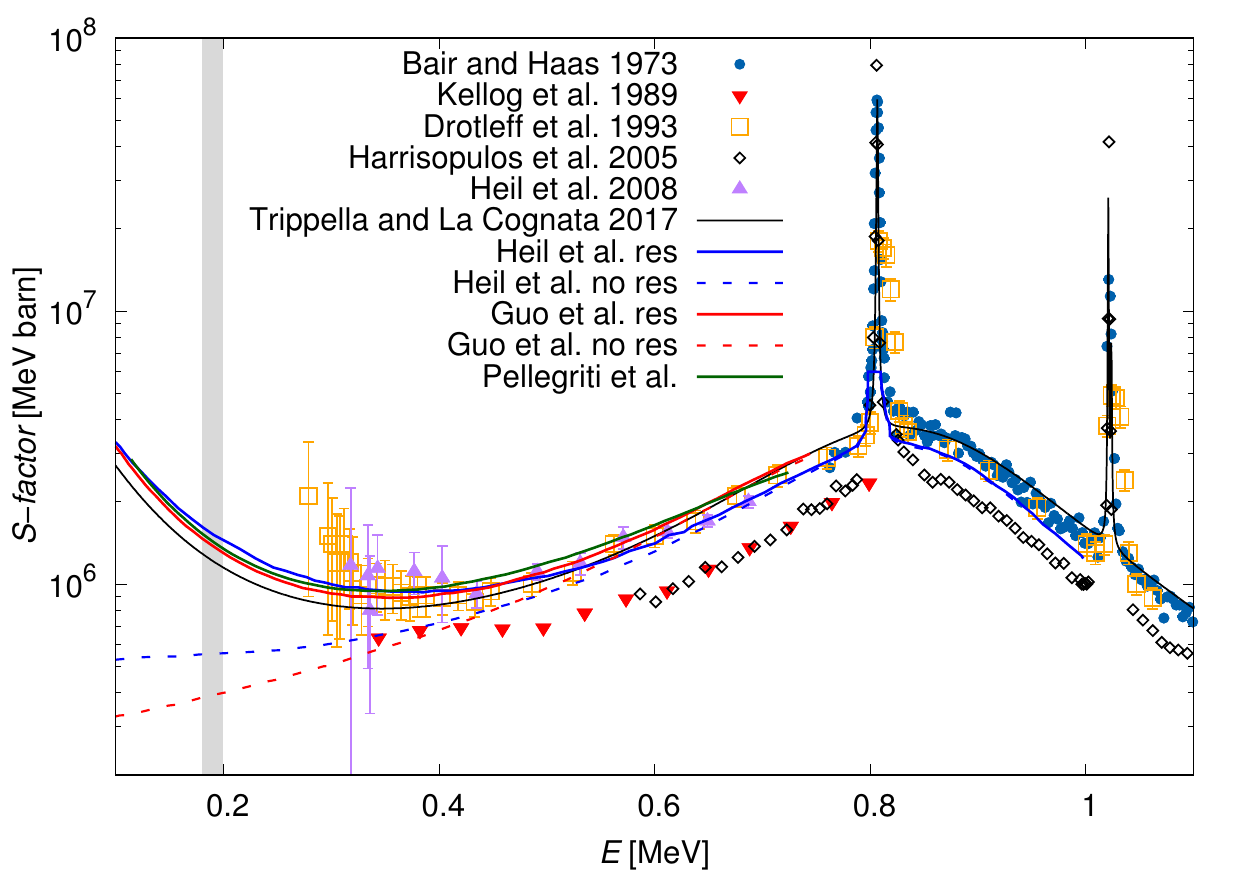}
\end{center}
\caption{The astrophysical S-factor for the \can\ reaction. No data exists in the energy region of astrophysical interest (grey area). The extrapolation to low energies (continuous red curve) is based on an R-matrix fit \cite{Heil08-PRC} assuming constructive interference with the $E = - 2.3$~keV sub-threshold state in $^{17}$O.
This extrapolation is rather similar to what has been obtained on the basis of the THM method \cite{LaCognata12-PRL,Trippella17-ApJ} and via transfer reactions \cite{Pellegriti08-PRC,Guo12-ApJ}.}
\label{fig:s-factor}
\end{figure}

In particular, the number of free neutrons in AGB stars determines the abundances of
elements heavier than iron and their elemental and isotopic ratios. As a consequence, the accurate and precise (at the level of 10\%) knowledge of the
$^{13}$C($\alpha$,n)$^{16}$O reaction is required at stellar temperatures in the range from 80 to 250 million K.
The current experimental situation, summarized in Fig.\ref{fig:s-factor}, does not fulfil such a requirement.
In particular,
the \can\ reaction ($Q = 2.216$~MeV) has been studied over a wide energy range by several direct measurements \cite{Davids68-NPA,Bair73-PRC,Ramstrom76-NPA,Kellogg89-BAPS,Drotleff93-ApJ,Harissopulos05-PRC,Heil08-PRC}.
From Fig.\ref{fig:s-factor} it is clear, first of all, that there exist no data from direct measurements close to the energy of astrophysical interest because of the severe limitations imposed by the high neutron background in surface laboratories. Second, the lowest energy data are affected by uncertainties that are too large to constrain extrapolations of higher energy data to astrophysical energies. Finally,  discrepancies exist between different data sets both in energy dependence and absolute values.

The extrapolation of experimental data to lower energies is further complicated by the unknown influence of three sub-threshold states and their possible interferences with higher energy resonances. In particular, the 1/2$^+$ state at $E = 6.356$~MeV in $^{17}$O, just 2.3~keV below the $\alpha$-particle threshold, is expected to provide the largest impact, but its contribution is still under discussion \cite{Avila15-PRC}.
The R-matrix extrapolation to low energies \cite{Heil08-PRC}  differs by up to a factor of 4 if a constructive interference with this
state is assumed or omitted. Only the Trojan horse method (THM) has provided a measurement of the subthreshold resonance \cite{LaCognata12-PRL,Trippella17-ApJ}.
The R-matrix calculated $S(E)$ factor obtained by using the THM resonance parameters is shown in Fig.\ref{fig:s-factor} together with the $S(E)$ factor indirectly obtained from transfer reactions \cite{Pellegriti08-PRC,Guo12-ApJ}.

LUNA has the potential to study \can\  in direct kinematics both with the 400 kV accelerator and then with the 3.5 MV one. This way it will be possible to cover a wide energy range, to address the issue of normalization discrepancies and to minimize overall statistical and systematic uncertainties, taking advantage of the 3 orders of magnitude suppression of the laboratory neutron background. Of course, also beam-induced background, mainly due to $^9$Be and $^{10,11}$B impurities in the solid $^{13}$C-enriched target
or along the beam line, has to be minimized.
In the energy region of interest, $E_\alpha = 0.3 - 1.4$~MeV neutrons are emitted with energies $E_{\rm n} = 2.0 - 3.5$~MeV thus requiring moderation before detection.
Given the low background, it may be possible to approach center of mass energies in the 200 keV range.

Finally, we point out that the same energy range of interest can be covered in inverse kinematics using a $^{13}$C beam with energies $E$ $= 0.9 - 4.5$~MeV.
A $^4$He gas target would be one possible solution, significantly changing the systematic uncertainties and background sources as compared to
the direct kinematics measurement.

\begin{figure}[t!]
    \centering
    \includegraphics[scale=1.4]{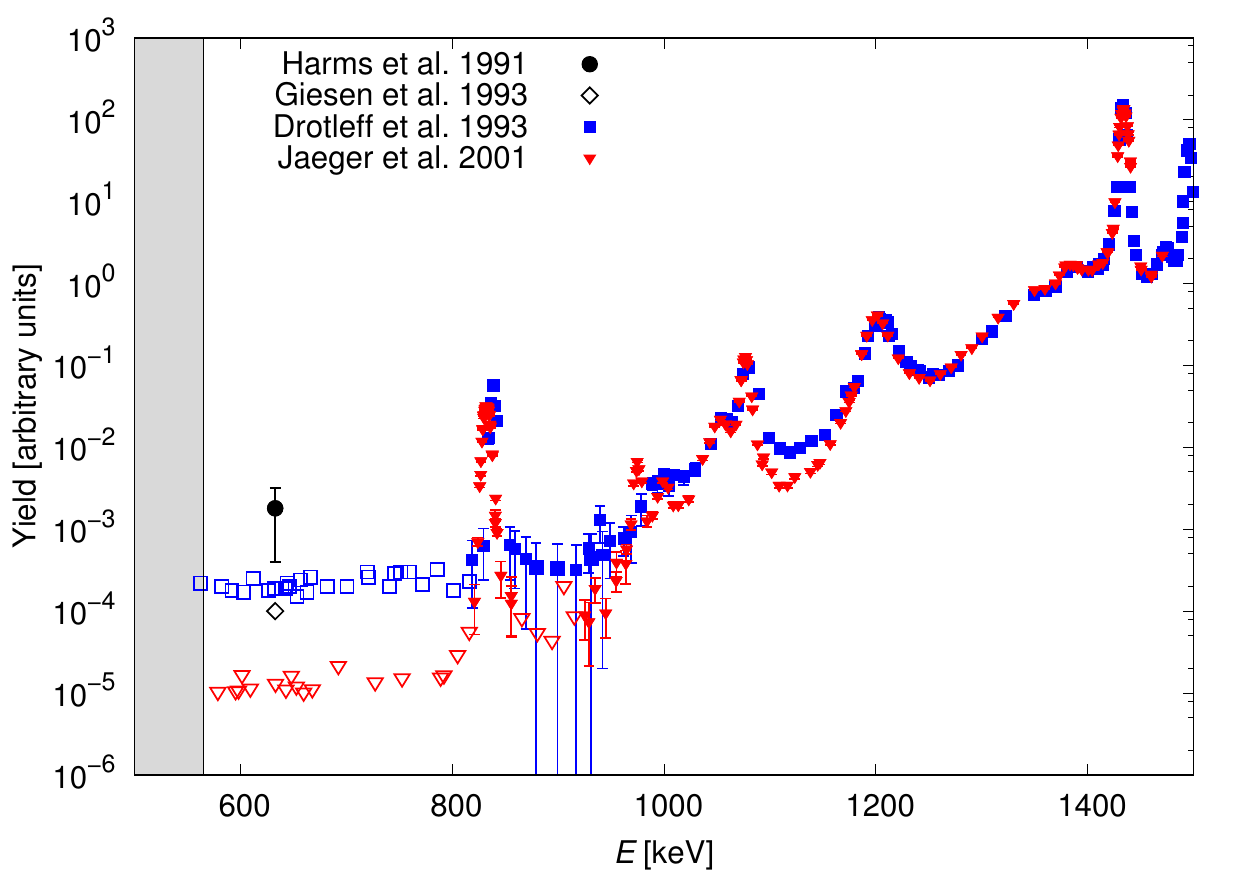}
    \caption{Excitation function of $^{22}$Ne($\alpha$,n)$^{25}$Mg from \cite{Jaeger01-PRL, Harms91-PRC, Drotleff93-ApJ, Giesen93-NPA}. Open points are upper limits. The grey area is kinematically excluded.}
    \label{fig:exfunc}
\end{figure}

\subsection{$^{22}$Ne($\alpha$,n)$^{25}$Mg}\label{subsec:ne22an}

$^{22}$Ne($\alpha$,n)$^{25}$Mg is switched on in AGB stars at the temperatures above $\sim$ 300
MK attained when the He-burning shell is activated. As a matter of fact, $^{22}$Ne($\alpha$,n)$^{25}$Mg
is the main source of neutrons in AGB stars of initial masses higher than roughly 4 solar masses. The abundances of the elements Rb and Zr
have been reported for these bright AGB stars \cite{Garcia06-Science,Garcia07-AA,Garcia09-ApJL}. The
observations qualitatively confirm the role of the $^{22}$Ne($\alpha$,n)$^{25}$Mg reaction as
the main neutron source, however, both models and observational uncertainties have hampered a
firm, quantitative comparison with model predictions \cite{vanRaai12-AA}. While observational
uncertainties are currently being addressed, model uncertainties both from
stellar physics and from the rate of $^{22}$Ne($\alpha$,n)$^{25}$Mg reaction remain troublesome.

Also for AGB stars of lower initial masses ($\sim$3 solar masses), where the
$^{13}$C($\alpha$,n)$^{16}$O reaction is the main neutron source, it has been shown
that the $^{22}$Ne($\alpha$,n)$^{25}$Mg neutron burst impacts the abundances of almost 200
nuclei \cite{Bisterzo15-MNRAS}.
This is due to their location nearby branching points on the path of neutron captures, where the s-process path
may move slightly off the valley of stability and which are
extremely sensitive to the high neutron densities.
Finally, the $^{22}$Ne($\alpha$,n)$^{25}$Mg reaction is also the main
neutron source for the s-process occurring during the hydrostatic burning of massive stars (with
initial mass greater than 10 solar masses) \cite{Pignatari10-ApJ}, which is responsible for the
cosmic production of the elements between the Fe peak and Sr.

The current knowledge of the
$^{22}$Ne($\alpha$,n)$^{25}$Mg reaction is incomplete and imprecise. Current estimates of the rate \cite{Bisterzo15-MNRAS}
are mostly based on experimental evaluations of the dominant resonance at 832 keV and provide
the rate with an uncertainty of 20-30\%, while less than 5\% is required for accurate model
predictions. Furthermore,
extrapolations to low energies may be affected by the
unknown influence of low-energy resonances just below the neutron threshold, casting doubts on
the accuracy of the values currently adopted in the stellar models.

The experimental panorama on the measurement of $^{22}$Ne($\alpha$,n)$^{25}$Mg is summarized in Fig.\ref{fig:exfunc} where the results of
the most sensitive experiment \cite{Jaeger01-PRL}   together with the ones of previous experiments \cite{Harms91-PRC,Drotleff93-ApJ,Giesen93-NPA} are shown.
In particular,
for $E_{\alpha}<800$ keV only an upper limit of about $\sigma < 10$~pb has been obtained. As a consequence, the reaction rates reported in literature  (and their uncertainties) strongly depend on theoretical assumptions related to the possible existence of unknown low-energy states.

It is possible to
study  $^{22}$Ne($\alpha$,n)$^{25}$Mg (Q=-478 keV)  with the $\alpha$ beam on a windowless gas target of enriched $^{22}$Ne surrounded by a neutron detector.
The most severe beam induced background is expected to arise from the $^{11}$B($\alpha$,n)$^{14}$N reaction, with $^{11}$B contained as impurity inside the components of the beam line.
It may be
feasible to have an accurate cross section measurement down to the center of mass energy
in the 600 keV range.
At lower energies it will be possible to exclude any resonance able to give a significant contribution to neutron production in AGB stars.

Finally, we point out that there are several others helium burning reactions which demand an underground study
such as ($\alpha$,$\gamma$) reactions on $^{2}$H, $^{14}$N, $^{15}$N, $^{17}$O and $^{18}$O.  In particular,
$^{12}$C($\alpha$,$\gamma$)$^{16}$O, one of the most important reactions of nuclear astrophysics
\cite{Iliadis15-Book}, will be the flagship of a second phase of LUNA MV.

\section{C burning}\label{sec:c12c12}

When He burning subsides, the lack of radiation pressure sends the core of the star into gravitational free fall, with the temperature increasing in response to the decrease in gravitational potential energy. This collapse is halted by one of two possible events: the temperature becomes sufficient to ignite the carbon in the star's core or electron degeneracy halts the contraction producing a carbon-oxygen white dwarf. Which of these paths a particular star follows is dependent on whether the ignition temperature for carbon fusion is reached ($\sim$ 5$\cdot$10$^8$ K) or, analogously, whether the critical mass for carbon ignition ($\sim$ 9M$\odot$)is exceeded prior to electron degeneracy \cite{Straniero16-JPCS}. Thus the end of helium burning marks a branching point in stellar evolution. Lower mass stars will become stable, electron-degenerate white dwarfs, while higher mass stars will enter the quiescent carbon burning phase of their evolution.
As a consequence, this limit separates the progenitors of white dwarfs, novae  and type Ia supernovae, from those of core-collapse supernovae, neutron stars, and stellar mass black holes. This mass limit also controls the estimations of the expected numbers of these objects in a given stellar population.

The rate of the $^{12}$C+$^{12}$C reaction, which is the trigger of the carbon burning, is a primary input to predict the behavior of a star at this branching point. Stellar models predict that quiescent carbon burning occurs for temperatures ranging between 0.5 and 1 GK, corresponding to center of mass energies between 0.9 and 3.4 MeV. However, the larger the $^{12}$C+$^{12}$C rate, the lower the temperature of the carbon burning. As a consequence, also the duration of the C burning is modified by a variation of the $^{12}$C+$^{12}$C rate. Carbon burning influences the energy generation and nucleosynthesis of massive stars. As a matter of fact, the two main channels of this reaction release protons and $\alpha$ particles in a rather hot environment, thus allowing a complex chain of reactions involving nuclei from C to Si. Some of these reactions, e.g. the $^{13}$C($\alpha$,n)$^{16}$O and the $^{22}$Ne($\alpha$,n)$^{25}$Mg, release neutrons and, in turn, activate the s-process which allows the production of heavy elements, as discussed in the previous section.

The $^{12}$C+$^{12}$C rate also affects the outcomes of type Ia supernovae \cite{Bravo11-AA}. It is believed that the reaction rate at temperatures as low as 0.15 GK (corresponding to an energy 0.7 MeV) may play a role in type Ia supernovae. We recall that type Ia supernovae play a fundamental role in cosmology, allowing the measurements of distances and of the expansion rates of high redshift galaxies.
The understanding of these different phenomena, of primary importance for astrophysics and cosmology, deserves a deeper experimental investigation of the $^{12}$C+$^{12}$C cross section.

\subsection{State of the art}

The $^{12}$C+$^{12}$C reaction, characterized by a Coulomb barrier of about 6.7 MeV, can proceed through different channels corresponding to the emission of a photon, a neutron, a proton, an $\alpha$ particle or even two $\alpha$ particles or a $^8$Be nucleus. Of these channels, the two more relevant are the emission of protons and $\alpha$ particles. The Q-value for proton emission is 2.24 MeV while that for $\alpha$ emission is 4.62 MeV. The proton and alpha channels can be measured either by detecting the charged particles or by revealing the gamma decay of the first excited state to the ground state of the $^{23}$Na or $^{20}$Ne residual nuclei, respectively. The energy of the two photons are 440 keV for the proton channel and 1634 keV for the alpha channel. In the gamma spectra, these energy peaks are severely doppler-shifted since the stopping time of the excited $^{23}$Na and $^{20}$Ne nuclei in the carbon target is comparable with their lifetimes.

\begin{figure}
\centering
\includegraphics[width=0.87\textwidth]{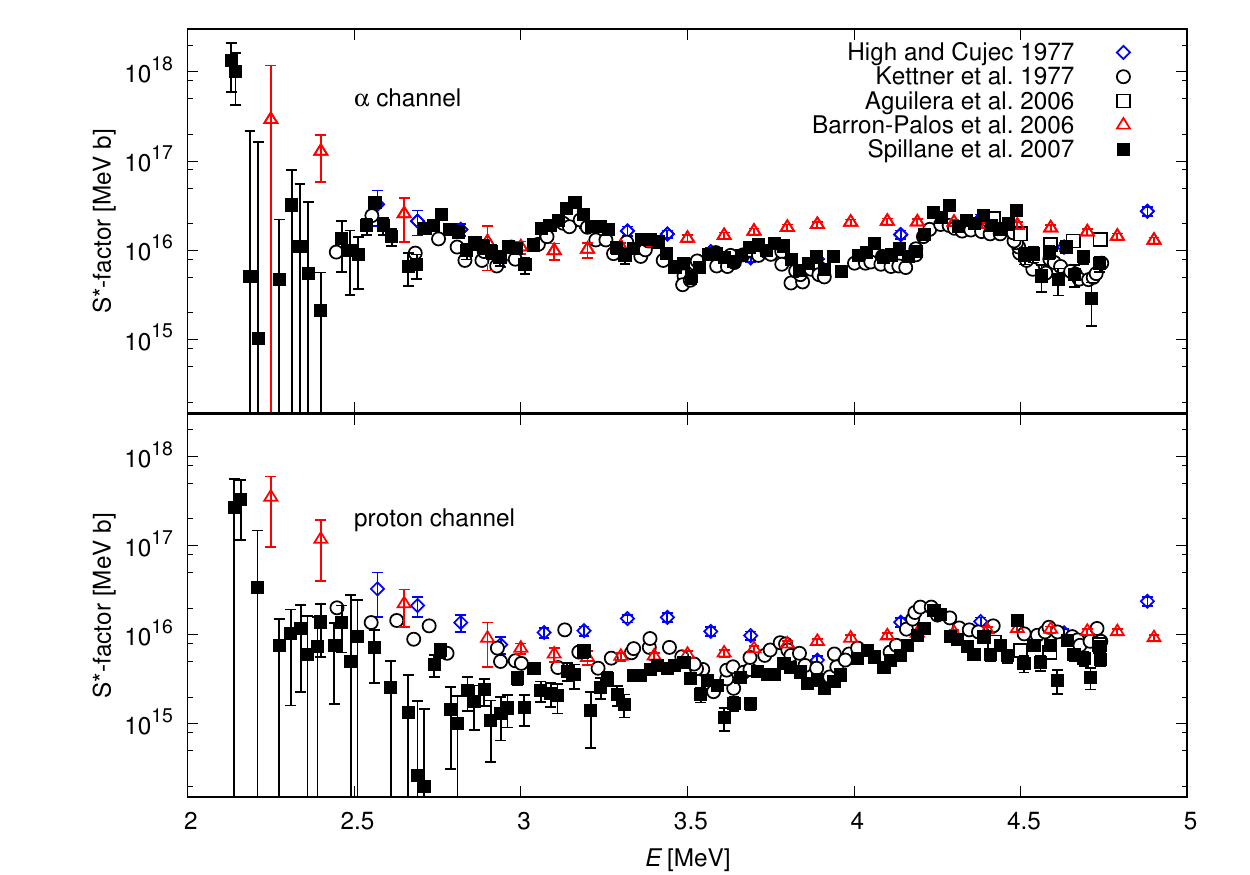}
\caption{ Modified astrophysical S-factor relative to the 440 keV transition (i.e. de-excitation of the first excited state of $^{23}$Na populated by the $^{12}$C($^{12}$C,$p$)$^{23}$Na reaction) and to the 1634 keV transition (i.e. de-excitation of the first excited state of $^{20}$Ne populated by the $^{12}$C($^{12}$C,$\alpha$)$^{20}$Ne reaction) }\label{fig:alessandra}
\end{figure}

Obviously, the gamma measurement cannot take into account the $\alpha_0$ and p$_0$ (particles with the full energy which leave the residual nucleus in the ground state) as well as the contributions from high energy states of the residual nuclei which de-excite directly to the ground state. Approximately, the decay of the first excited state to the ground state accounts for 50\% of the total cross section. So far, many different experiments attempted to measure the $^{12}$C+$^{12}$C reaction using one of the two above described techniques or both. The first experiment dates back to 1960 \cite{Almqvist60-PRL}  while the most recent are that of Spillane et al. \cite{Spillane07-PRL}  and that presently  ongoing at the CIRCE accelerator in Caserta, Italy \cite{Morales15-JPCS}. A summary of the results \cite{High77,Kettner77-PRL,Aguilera06,Barron06-NPA,Spillane07-PRL} in terms of the modified astrophysical S factor (which includes the first order correction to the penetrability since the charge of the two interacting nuclei are rather high) as a function of the energy is presented in Fig.\ref{fig:alessandra}.

The lowest energy at which the cross section is measured is 2.1 MeV \cite{Spillane07-PRL}. The general structure is characterized by the presence of several resonances superimposed on a flat background. The resonances have a typical width of 10 keV and are spaced by 300-500 keV. The lowest energy resonance observed (E = 2.14 MeV \cite{Spillane07-PRL}) has a quite clear signature in the alpha channel but is unresolved in the proton channel due to the large uncertainties of the data at these energies. Nevertheless, it is characterized by a relative large strength and its impact on the reaction rate is very relevant.

The measurement of the $^{12}$C+$^{12}$C reaction is affected by beam induced as well as natural background issues. The former are primarily due to impurities in the carbon target, since the $^{12}$C beam purity achieved so far is extremely high, with contaminations less than one part in 10$^{12}$. Beam induced background is significant only if the reaction producing it has a cross section much greater than $^{12}$C + $^{12}$C. Using the Coulomb barrier criterium, only impurities due to  H, He, Li, Be, B and C should be considered. The cross sections of  $^{12}$C + $^{13}$C and $^{12}$C + $^{12}$C have been shown to be comparable at energies near and below the Coulomb barrier. As the isotopic ratio of $^{13}$C/$^{12}$C is 0.01, $^{12}$C + $^{13}$C should not contribute meaningfully to the measured yield.
Backgrounds due to isotopes of He, Li and B were investigated by Spillane et al. and found to be negligible. On the contrary, a small Be contamination was not completely ruled out \cite{Spillane06-PhD}. However, the most prominent background is due to hydrogen and deuterium because of their ease of forming bonds with carbon. These ions are also deposited on the surface of carbon targets from the vacuum rest gas during measurements.

For what concerns gamma measurements, previous experimental works have identified the $\gamma$-rays  from the $^{2}$H($^{12}$C,p$_1\gamma$)$^{13}$C and $^{1}$H($^{12}$C,$\gamma$)$^{13}$N reactions as primary sources of beam induced background: they emit $\gamma$-rays at 3.09 MeV and 2.36 MeV, respectively.  At low beam energies, the Compton background of these peaks could completely dominate the carbon fusion $\gamma$-ray peaks, as evidenced in \cite{Kettner77-PRL} and \cite{Barron06-NPA}. For what concerns particle measurements, if the particle detectors are placed at backward angles, it is kinematically impossible to find protons in the carbon fusion region of interest (ROI) from nuclear reactions of $^{12}$C with $^1$H or $^{2}$H:  the $^1$H($^{12}$C,p)$^{12}$C reaction will not produce protons at backwards angles, while the $^2$H($^{12}$C,p)$^{13}$C reaction produces protons with significantly lower energies than the carbon ROI. Unfortunately, a two-step process is possible in the presence of deuterium \cite{Zickefoose10-PoS}. When deuterium is elastically scattered by carbon at forward angles followed by the $^{12}$C(d,p)$^{13}$C reaction in the target, the kinematics allows protons to be produced directly in the relevant ROI. Thus it is of great importance to produce extremely pure targets. Furthermore, the rest gas must be monitored and controlled to reduce hydrogen isotope contamination. Moreover, periodic analysis of deuterium contamination is necessary for accurate background subtraction.
Spillane et al. \cite{Spillane07-PRL} found a method to  mitigate the hydrogen and deuterium content: they placed the target in a chamber under vacuum and exposed it to an intense  $^{12}$C beam bombardment without any cooling. By heating up the target for 20 minutes at  700$^{\circ}$C the contamination was reduced to a negligible level.
In order to fulfill this procedure, thick targets are required to stand the intense beam bombardment during the heating.

With the reduction or elimination of hydrogen and deuterium from the target, the primary background  in the $\gamma$-ray spectrum derives from naturally occurring sources, primarily ubiquitous natural radioisotopes. This background is negligible at higher energies, but becomes significant below 3.0 MeV.
In the measurement of Spillane et al. \cite{Spillane07-PRL}, the HPGe detector was surrounded by a 15 cm thick lead shield allowing a reduction of the natural background by a factor of 400 near E$_\gamma$= 1.6 MeV. As a matter of fact, in a laboratory at the Earth's surface, the shielding efficiency cannot be increased by further adding any more shield since the cosmic muons would interact with the added material, creating more background. Of course, this problem is dramatically reduced in the Gran Sasso underground laboratory where the rock overburden reduces the muon component of the cosmic background by a factor of 10$^{6}$, as already underlined. Indeed, in the case of the $^{3}$He($^{4}$He,$\gamma$)$^{7}$Be measurement \cite{Costantini08-NPA}, with a proper massive shielding of 0.3 m$^{3}$ of copper and lead, surrounded by an anti-radon envelope of plexiglas flushed with N$_{2}$ gas, a background suppression of 5 orders of magnitude was reached for $\gamma$ rays below 2 MeV with
respect to a background spectrum measured underground with no shielding \cite{Caciolli09-EPJA}.  Therefore, a deep underground measurement represents a unique possibility to reach the low energy domain of the $^{12}$C~ +~ $^{12}$C reaction. The advantage of an underground measurement is less evident in the case of particle detection, even if a recent measurement performed at the LUNA 400 kV accelerator proved that revealing low energy alpha particles is easier in a deep underground laboratory than overground \cite{Bruno15-EPJA}.

$^{12}$C+$^{12}$C is now supposed to be the flagship reaction of the first 5 year scientific program with the new 3.5 MV accelerator. In the energy range 0.5-3.5 MeV, the expected intensity is $>$100 $\mu$A for the $^{12}$C$^{+}$ beam. Since the energy in the center of mass system is exactly one-half of the beam energy, the measurement can be performed with the more intense single charge state beam for E$_{cm}$ $\leq$ 1750 keV, approximately. For higher energies, only the $^{12}$C$^{++}$ beam can play the game. The beam will impinge on a solid $^{12}$C target of natural composition with the as low as possible contamination due to hydrogen isotopes.

As for the detection system, a high efficiency and ultra low intrinsic background HPGe detector, as the one now in use at the LUNA 400 kV accelerator \cite{Caciolli09-EPJA}, complemented with  silicon detectors (or telescopes) at backward angles may allow the measurement of both the $\gamma$-rays and the particles (protons and alphas).
The deep underground location of the LUNA MV accelerator and its capability of producing an intense carbon beam offer a unique opportunity to perform such a measurement entering for the first time the low energy region below 2 MeV.

\begin{figure}[t!]
\begin{center}
\includegraphics[width=1.03\textwidth]{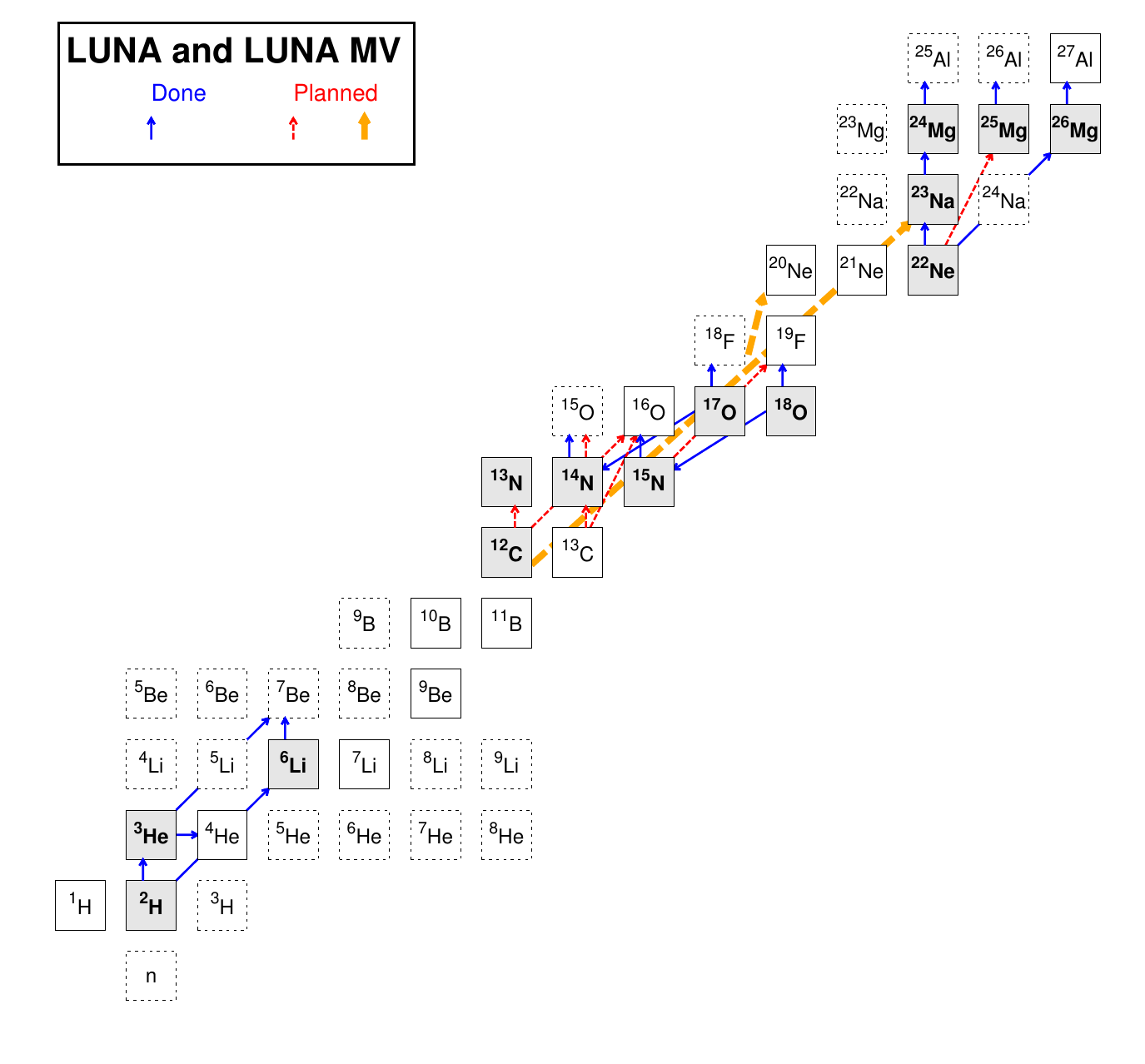}
\end{center}
\caption{The LUNA reactions in the nuclide chart. The yellow lines highlight the two carbon burning reactions which are supposed to be the flagship of LUNA MV first phase:
$^{12}$C($^{12}$C,p)$^{23}$Na and $^{12}$C($^{12}$C,$\alpha$)$^{20}$Ne.
}
\label{fig:nuclides}
\end{figure}

\section{Conclusions}\label{sec:Conclusion}

In 1991 LUNA started underground nuclear astrophysics in the core of Gran Sasso, below 1400 meters of
dolomite rock. The extremely low background has allowed for nuclear physics experiments with very small count rate, down to a few events per year. The important reactions responsible for the hydrogen burning in the Sun have been studied for the first time down to the relevant energies, providing fundamental contributions for the prediction of the solar neutrino spectrum.

At the end of the solar phase, LUNA started a rich program, still going on, devoted to the study of BBN and of the synthesis of the light
elements through the CNO, Ne-Na and Mg-Al cycles. Thanks to the measurements already done, it has been possible to reduce the error on the prediction of oxygen and fluorine isotopes
from AGB stars and classical novae. In particular, the production site of a star grain population has been firmly identified thanks to a new isotopic signature coming from LUNA results.
In addition,  it has been proven that the cross sections of $^{2}$H($\alpha$,$\gamma$)$^{6}$Li is not the reason for the primordial $^{6}$Li  problem.
As a summary of the LUNA activities, Fig.\ref{fig:nuclides} shows the nuclide chart from hydrogen to aluminium with the past, present and future of LUNA.

In particular, starting in 2019 LUNA will enter a new phase: the study of helium and carbon burning with a 3.5 MV accelerator able to deliver high current beams of
hydrogen, helium and carbon (also double ionized).
The program of the first five years of running time is now supposed to  be focused on the study of
$^{12}$C+$^{12}$C, the reaction which
is switching on the carbon burning. In particular, its rate determines the evolution of a massive star up to a
slowly cooling white dwarf or up to a core-collapse supernova. The other two key reactions which will be studied during this phase are
$^{13}$C($\alpha$,n)$^{16}$O and $^{22}$Ne($\alpha$,n)$^{25}$Mg, responsible for generating free neutrons inside stars. These neutrons give then rise to the s-process:
the synthesis of about half of the chemical elements beyond the iron peak in the Universe.

\section*{Acknowledgments}

It is a pleasure to thank our Colleagues of LUNA, without whom the results reviewed in this paper would not have been achieved.
In particular, we thank Alessandra Guglielmetti for her contributions to the carbon burning section and for a critical reading of the manuscript, and Maria Lugaro for several enlightening discussions on the s-process.


\end{document}